\documentclass[prx,reprint,twocolumn,showpacs,superscriptaddress]{revtex4-1}
\usepackage{amsmath,amssymb,bm,mathrsfs,graphicx, braket, times,enumerate}
\usepackage[colorlinks=true,citecolor=blue,linkcolor=blue]{hyperref}
\usepackage{longtable}
\usepackage{multirow}
\usepackage[all,cmtip]{xy}
\usepackage[normalem]{ulem}
\usepackage[usenames,dvipsnames]{color}

\renewcommand{\vec}[1]{\bm{#1}}

\newcommand{\sg}[1]{{\bf\em{#1}}}

\begin{document}

\title{Lattice Homotopy Constraints on Phases of Quantum Magnets}

\author{Hoi Chun Po}
\affiliation{Department of Physics, University of California, Berkeley, CA 94720, USA}
\affiliation{Department of Physics, Harvard University, Cambridge MA 02138, USA}

\author{Haruki Watanabe}
\affiliation{Department of Applied Physics, University of Tokyo, Tokyo 113-8656, Japan.}

\author{Chao-Ming Jian} 
\affiliation{Station Q, Microsoft Research, Santa Barbara, California, 93106, USA.}
\affiliation{Kavli Institute for Theoretical Physics, University of California, Santa Barbara, California, 93106, USA}

\author{Michael P. Zaletel}
\affiliation{Department of Physics, Princeton University, Princeton, NJ 08544, USA.}

\begin{abstract}
The Lieb-Schultz-Mattis (LSM) theorem and its extensions forbid trivial phases from arising in certain quantum magnets. Constraining infrared behavior with the ultraviolet data encoded in the microscopic lattice of spins, these theorems tie the absence of spontaneous symmetry breaking to the emergence of exotic phases like quantum spin liquids. In this work, we take a new topological perspective on these theorems, by arguing they originate from an obstruction to ``trivializing'' the lattice under smooth, symmetric deformations, which we call the ``lattice homotopy problem.'' We  conjecture that all LSM-like theorems for quantum magnets (many previously-unknown) can be understood from lattice homotopy, which automatically incorporates the full spatial symmetry group of the lattice, including all its point-group symmetries. 
One consequence is that any spin-symmetric magnet with a half-integer moment on a site with even-order rotational symmetry must be a spin liquid. To substantiate the claim, we prove the conjecture in two dimensions for some physically relevant settings.
\end{abstract}
 
\maketitle

%\section{Introduction}

Quantum magnets arise naturally in Mott insulators, where strong Coulomb repulsion freezes the position of electrons and leaves behind their spin degrees of freedom.  With strong frustration, quantum fluctuations can suppress spin ordering and lead to symmetric, quantum-entangled phases of matter that survive down to zero temperature.  Quantum spin liquids, the spin analogues of fractional quantum Hall states, represent one of the most sought-after phases arising in this context \cite{BalentsSL}. They possess intrinsic topological order with emergent fractionalized excitations, which have been proposed as a useful resource for robust quantum computation \cite{FreedmanTopoComp,RevModPhys.80.1083}.

Detecting whether a quantum magnet is a spin liquid, a many-body problem, is notoriously hard.
Conventionally, the absence of symmetry breaking is regarded as an indicator for spin-liquid physics \cite{RevModPhys.80.1083, PhysRevLett.91.107001}. However, a symmetric quantum magnet could also be in a symmetry-protected topological (SPT) phase, like the spin-1 Haldane chain \cite{Haldane, AKLT} and its generalizations \cite{PhysRevB.84.235141}, which does not support fractionalized excitations despite a nontrivial degree of entanglement. Conceptually, there is a sharp distinction between these phases: spin liquids are long-range entangled (LRE), and are necessarily either gapless or topologically ordered, while SPT phases are only short-range entangled (SRE). 
Experimentally, however, such distinction is subtle and one must rely on additional criteria to rule out \emph{all} symmetric SRE (sym-SRE) phases before claiming discovery of a spin liquid.

Fortunately, it is possible to rule out all sym-SRE phases in certain quantum magnets on purely theoretical grounds. This line of reasoning was pioneered by Lieb, Schultz, and Mattis (LSM), who proved that any one-dimensional quantum magnet with both lattice-translation and spin-rotation symmetries cannot be sym-SRE if each unit cell contains a half-integral total spin \cite{Lieb1961}. Multiple generalizations of the LSM theorem have since been made, covering systems in higher dimensions and with less stringent physical assumptions
\cite{Affleck1986, AffleckPRB,  YOA-PRL, Oshikawa2000,Hastings2004,Xie2011,Sid2013,LSM_Roy,PNAS}. We will collectively refer to these results as ``LSM-like theorems.'' The common denominator of these generalizations is a constraint between the microscopic details of the system, specifically the lattice and the symmetry-transformation properties of the sites' Hilbert spaces, and the degree of ground-state degeneracy. Since any sym-SRE phase is expected to have a gapped, unique ground state on any thermodynamically large space without defects or boundaries, all sym-SRE phases are ruled out whenever the degeneracy is constrained to be nontrivial.
In this sense, the LSM-like theorems are ``no-gos" for sym-SRE phases.

One important direction for generalization is to make fuller use of the spatial symmetries of the system. This was partially addressed in Refs.~\cite{Sid2013,LSM_Roy,PNAS}, which showed that combinations of nonsymmorphic symmetries like glides and screws, being ``fractions" of the lattice translations,  can lead to stronger no-gos. Ideally, to expose the strongest constraints one would attempt to utilize \emph{all} spatial symmetries of the problem. However, the nonsymmorphic generalizations in Refs.~\cite{Sid2013, LSM_Roy, PNAS} ignore all point-group symmetries (e.g., rotations), which fix at least one point in space. New techniques are required for the desired extension.

In this work, we address the problem of incorporating all spatial symmetries in deriving stronger LSM-like no-gos, which are operative even when all  earlier theorems are not.
We will rely on two key insights: First, the presence of a no-go should be insensitive to a smooth, symmetric deformation of the underlying lattice. We will refer to the study of such deformations as the ``lattice homotopy problem;" second, there is a strong sense of locality in sym-SRE phases due to the limited range of entanglement, and therefore, compared to more exotic phases like spin liquids, they respond in a more conventional manner when fluxes are inserted into the system. 
Combining these observations, we conjecture that a quantum magnet which is nontrivial under lattice homotopy is obstructed from being sym-SRE.

In the following, we will elaborate on the conjecture, which encompasses all earlier LSM-like theorems for quantum magnets, and then offer a physical argument for its proof restricting to 2D systems with an internal symmetry group $G$ being either finite Abelian or ${\rm SO}(3)$.
As an example, we will show that sym-SRE phases are forbidden whenever a half-integer spin, carrying a projective representation of ${\rm SO}(3)$, sits at an even-order rotation center. 
Intuitively, this is because any symmetric deformation brings in an even number of spins, which cannot screen the half-integer moment at the center.

\begin{figure}
\begin{center}
{\includegraphics[width=0.48 \textwidth]{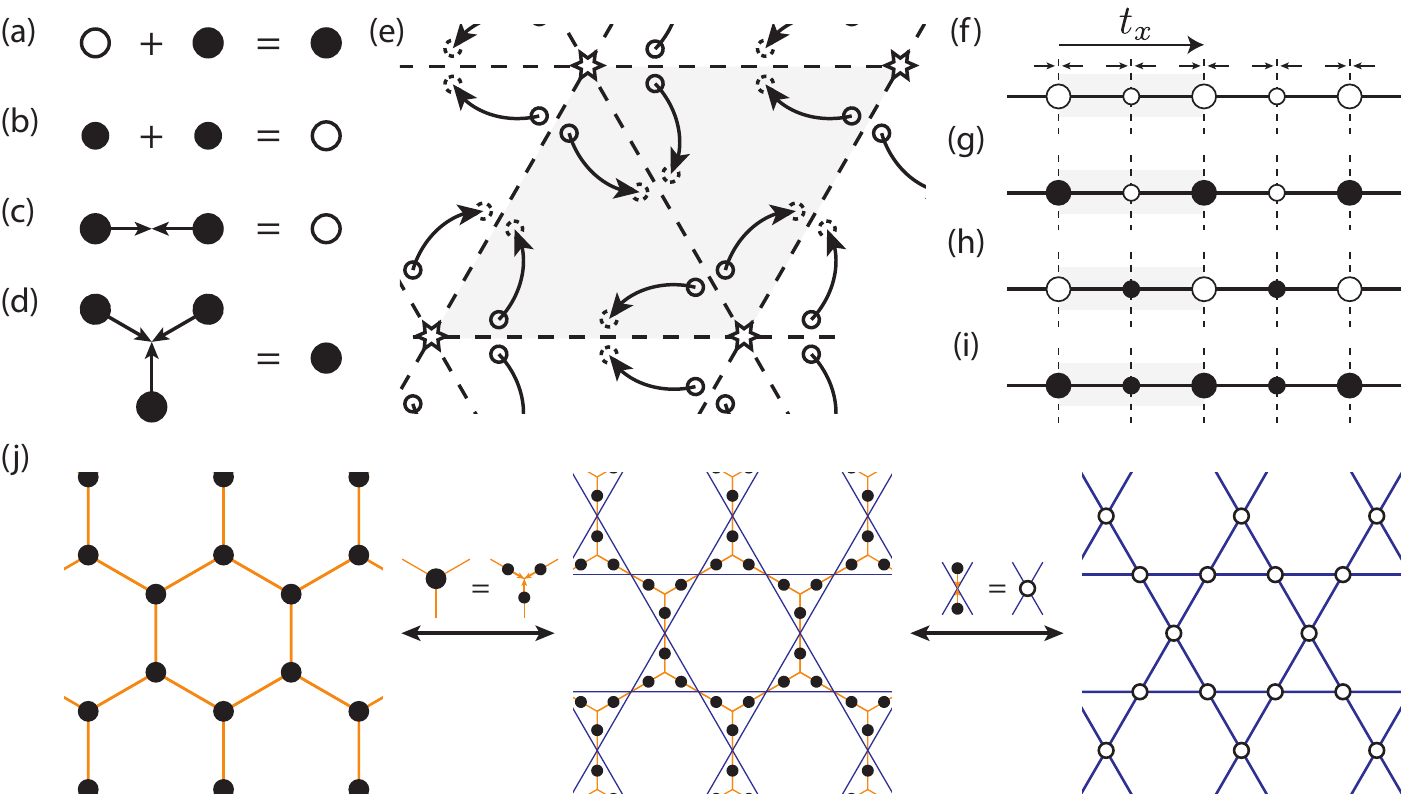}} 
\caption{{\bf Lattice homotopy.} (a-d) Representations of the rotation group ${\rm SO}(3)$ fuse following a $\mathbb Z_2$ rule. Open and filled circles respectively denote the representations of integer and half-integer spins. 
(e) A smooth deformation of a lattice (circles) symmetric under mirror planes (dashed lines) and three-fold rotations (about the stars).
(f-i) There are two inequivalent sites (big and small circles) in each unit cell (shaded) of a mirror-symmetric 1D lattice. Under lattice homotopy, there are four distinct lattice classes.
(j) Assuming the symmetries of (e), a honeycomb lattice of half-integer spins is equivalent to a kagome lattice of integer spins, as demonstrated by the depicted smooth deformation.
\label{fig:Latt_Homo}}
\end{center}
\end{figure}

{\it Statement of the conjecture.}--
Consider a quantum magnet with Hamiltonian $\hat H$ defined on a lattice $\Lambda$. 
For simplicity, we will first assume $\hat H$ is symmetric under the group $G = {\rm SO}(3)$ of spin rotations, and later discuss how the ideas apply to more general on-site symmetry groups.
We are interested in whether $\hat H$ can be in a sym-SRE phase. As demonstrated by the LSM-like theorems, the microscopic data encoded in $\Lambda$ may present an obstruction. The key ingredient in our argument will be the spatial distribution of half-integer vs.\ integer spins.
Therefore, as far as obstructions are concerned, we view $\Lambda$ as a lattice of black and white circles,  denoting half-integer and integer spins respectively (Fig.~\ref{fig:Latt_Homo}).
No obstruction is expected on a lattice composed only of integer spins, and we say such lattices are ``trivial." 
In addition, the presence of obstructions should  be insensitive to a smooth deformation of the lattice, provided that the deformation respects all spatial symmetries (Fig.~\ref{fig:Latt_Homo}e).
This motivates the following conjecture:\\

\noindent{\bf Conjecture}: \emph{A sym-SRE phase is possible only when $\Lambda$ is smoothly deformable to a trivial lattice.}\\

Let us make  precise what is meant by a ``smooth deformation." We suppose the magnet is symmetric under a space group $\mathcal S$. By deformation, we refer first to a collective, $\mathcal S$-symmetric movement of sites.
Second, when sites collide they ``fuse;" since we only keep track of the integer vs.\ half-integer nature of the sites, the fusion follows a $\mathbb{Z}_2$-rule (Fig.~\ref{fig:Latt_Homo} a-d). In this process, an even number of half-integer sites may annihilate.
Generally, when a collection of sites are symmetrically collapsed at a point, the number of sites involved is determined by the degree of the point-group symmetry.
We also allow the inverse of fusion, in which half-integer spins are created in pairs.

A sequence of such deformations defines an equivalence relation between lattices, and we refer to the enumeration of the resulting equivalence classes $[\Lambda]$ as the ``lattice homotopy" problem.
$\{ [\Lambda]\}$ naturally forms an Abelian group under stacking, with the empty (trivial)  lattice the identity element.
The conjecture is that a sym-SRE obstruction is present whenever a lattice belongs to a nontrivial class.
We  note that all the previously-known LSM-like theorems feature nontrivial lattices \cite{Lieb1961,Affleck1986, AffleckPRB,  YOA-PRL, Oshikawa2000,Hastings2004,Xie2011,Sid2013,LSM_Roy, PNAS}.

Thanks to its geometrical nature, lattice homotopy can often be computed in an intuitive manner. For instance, consider a 1D translation and mirror symmetric spin chain. Spins at generic positions can be smoothly brought to the mirror planes, where they will annihilate pairwise.
Since this cannot change the color on the mirror plane, the only lattice invariant is the color at the two inequivalent mirror planes in a unit cell, giving a $ [\Lambda] \in  \mathbb Z_2 \times \mathbb Z_2$ classification (Fig.~\ref{fig:Latt_Homo}f-i). In fact, a no-go for the three nontrivial elements was already proven in Ref.~\cite{PNAS}. Together with the original LSM theorem invoking only translations, this proves the conjecture for the two 1D space groups.

As another example, a square lattice of spin-1/2 moments is nontrivial, but that of spin-1 is trivial. This is consistent with the known LSM-like theorems for the former \cite{Affleck1986, AffleckPRB, YOA-PRL, Oshikawa2000,Hastings2004}, and the existence of sym-SRE phases for the latter~\cite{JianS1}.  A more intriguing example is a honeycomb lattice of half-integer spins, which is symmetric under both three-fold rotations and mirrors. As shown in Fig.~\ref{fig:Latt_Homo}j, the lattice is smoothly deformable to a kagome lattice of integer spins, and therefore belongs to the trivial class. Interestingly, this picture is consistent with a recent construction of sym-SRE wave-functions \cite{JianS1,KimHoneycomb}.

It is conceptually revealing to generalize the internal symmetry group $G$ in the discussion above beyond ${\rm SO}(3)$ spin-rotations.
We assume that the total symmetry group is a direct product of the internal and space group symmetries, $G \times \mathcal{S}$, where $\mathcal{S}$ acts by permuting the sites. (The case with ``spin-orbit coupling" is an interesting future direction.)
The role of ``half-integer" vs.\ ``integer" spin is now played by the Abelian group of distinct projective representations of $G$, $\mathcal{H}^2 \left[ G, U(1) \right]$. 
The $\mathbb{Z}_2$-fusion of spins generalizes to group multiplication in $\mathcal{H}^2 \left[ G, U(1) \right]$, and the above conjecture naturally carries over to this more general setting.

The resulting group of lattice homotopy classes depends on $G$. For instance, suppose $G$ is such that the projective representations have $\mathbb{Z}_3$ fusion, and consider again the 1D lattice with reflection symmetry. If two copies of a projective representation $[\omega]$ approach a mirror plane, they do \emph{not} annihilate, since $[\omega]^2 = [\omega]^{-1}$ in $\mathbb{Z}_3$.
Consequently, the projective representation on a mirror plane is not conserved, and the lattice homotopy classification collapses down to $\mathbb Z_3$.

Computing the lattice classification can be automated by a reduction to the properties of high-symmetry points (Wyckoff positions). We relegate details Appendix \ref{app:latticehomotopy}.
In Table~\ref{tab}, we tabulate the lattice classification results for all 2D space-groups. 
Here, we present the case relevant to spins, $\mathcal{H}^2 \left[ G, U(1) \right] = \mathbb{Z}_2$ -- the general form, which extends readily to  any finite Abelian $\mathcal{H}^2 \left[ G, U(1) \right]$, is tabulated in Supplementary Table I.

\begin{table}
\begin{center}
\caption{The lattice homotopy classification for the 17 wallpaper groups, assuming $\mathbb{Z}_2$ projective representations, as in the case for spin-rotation invariant quantum magnets. \label{tab}}
\begin{tabular}{cl}\hline\hline
Lattice homotopy & Wallpaper group No.~\cite{ITC}\hspace{10pt}\mbox{} \\ \hline
$\mathbb{Z}_2$ 			& 1, 4, 5, 13, 14, 15\\
$\left( \mathbb{Z}_2 \right)^2$ 	& 3, 8, 12, 16, 17\\
$\left( \mathbb{Z}_2 \right)^3$ 	& 7, 9, 10, 11\\
$\left( \mathbb{Z}_2 \right)^4$ 	& 2, 6 \\
\hline\hline
\end{tabular}
\end{center}
\end{table}

{\it Proof of conjecture in 2D.}--
We now sketch a physical argument supporting the conjecture for quantum magnets symmetric under any of the 17 2D space groups, assuming $G = {\rm SO}(3)$ or is finite Abelian.
The logic proceeds by first deriving three concrete conditions on $\Lambda$, each implying  a no-go for sym-SRE phases:
\begin{itemize}
\item{Bieberbach no-go.} A ``fundamental domain" $D$ is a region which tiles the plane under the action of translation and glide symmetries. If the total projective representation in $D$ is nontrivial, $[\omega]_D = \prod_{\vec r \in D} [\omega]_{\vec r} \neq 1$, then a sym-SRE phase is forbidden \cite{PNAS}.

\item{Mirror no-go.} Let $\ell$ be a mirror-line parallel to a translation $T_\parallel$. We define the projective representation per unit length of $\ell$,  $[\omega]_\ell = \prod_{\vec r \in \ell'} [\omega]_{\vec r}$, by letting the product runs over a unit-length interval $\ell'$ of $\ell$ as defined by $T_{\parallel}$. If $[\omega]_\ell$  does not have a ``square-root," i.e., if no $\zeta\in\mathbb{Z}_n$ satisfies $[\omega]_\ell = \zeta^2$, then a sym-SRE phase is forbidden \cite{PNAS}.

\item{Rotation no-go.} Let $\vec r$ be a site with rotational point-group symmetry $C_m$ and projective representation $[\omega]_{\vec r}$. If $[\omega]_{\vec r}$ does not have an ``$m$-th root", i.e., if no $\zeta\in\mathbb{Z}_n$ satisfies $[\omega]_{\vec r} = \zeta^m$, then a sym-SRE phase is forbidden.
\end{itemize}
We then show that these no-gos forbid a sym-SRE phase in a 2D lattice $\Lambda$ if and only if $[\Lambda]\neq 1$. 
Both the Bieberbach and mirror no-gos were derived in an earlier work \cite{PNAS}, so here we focus on illustrating the key ideas behind the derivation of the ``rotation no-go" -- the key missing piece for establishing the conjecture in 2D -- with further details  given in Appendix \ref{app:Proof}.

{\it Derivation of the rotation no-go.}--
For simplicity, we will illustrate the ideas using systems symmetric under $G={\rm SO}(3)$ and $C_2$ rotation.
Roughly speaking, we will modify the Hamiltonian by inserting a pair of $C_2$-related spin fluxes, and show that when a half-integer  moment lies on a $C_2$-invariant point, the system has a symmetry-protected degeneracy.
We will then argue that, despite the presence of fluxes, such degeneracy remains impossible in sym-SRE phases, and thereby arriving at a no-go.

We begin with the following observation: While a sym-SRE phase has a gapped, unique ground state on $\mathbb R^d$, it may possess symmetry-protected ground-state degeneracy in the presence of defects or boundaries (a notable example being the edge states of the AKLT chain).
In contrast to LRE phases, however, the degeneracies in a sym-SRE phase should be ``localized" to the defect regions (for example, each edge of the AKLT chain carries an independent two-fold degeneracy).
Physically, this arises because a sym-SRE phase can only respond to local data, defined with respect to the correlation length $\xi$, so it should not be possible to ``share'' a degeneracy between two distant defect regions (note we are only considering bosonic models; certain fermionic SPTs violate this assumption \cite{Fidkowski2016}).

To formalize this intuition we introduce the notion of ``\emph{degeneracy localization}'' (Appendix \ref{app:DegLoc}).
A ``defect region" is a region in which the Hamiltonian is not local-unitarily equivalent to the Hamiltonian of the bulk \cite{PNAS} (examples could include an impurity spin, dislocation, or external flux), and we let $\{R^{(i)}:i=1,\dots,N_D\}$ be a collection of defect regions of finite extent, which are separated from each other on distances $r\gg \xi$.
We say the system exhibits \emph{degeneracy localization} if each $R^{(i)}$ can be modeled as an emergent $d_i$-dimensional, degenerate, degree of freedom, so that the total ground-state subspace  $\mathcal H_{\rm GS}'$ is $\left(\prod_{i=1}^{N_D}d_i\right)$-dimensional.
This implies that if $\hat U$ is a local operator taking the ground-state subspace $\mathcal H_{\rm GS}'$  into itself (e.g., a symmetry), then its projection into $\mathcal H_{\rm GS}'$ can be ``factorized" as $\hat U|_{\rm GS} = \bigotimes_{i=1}^{N_D} U^{(i)} + \mathcal O(e^{-r/\xi})$ for some $d_i$-dimensional matrix $U^{(i)}$ acting only on the degeneracy localized at the region $R^{(i)}$. In other words,  degeneracy localization passes the locality structure from the full Hilbert space onto $\mathcal H_{\rm GS}'$. 
The discussed intuition about sym-SRE phases can then be summarized by the following physical assumption: \emph{A bosonic sym-SRE phase exhibits degeneracy localization.}

We now use this assumption to prove the $C_2$-rotation no-go with $G={\rm SO}(3)$.
Recall that a (projective) representation of ${\rm SO}(3)$ is classified by $[ \omega]_{\vec r} \in \mathbb Z_2 = \{1,-1 \}$, which encodes the phase factor for the commutator of two orthogonal $\pi$-rotations at site $\vec r$, say
$
\hat X_{\vec r} \hat Z_{\vec r}  =[ \omega]_{\vec r}\hat Z_{\vec r}\hat X_{\vec r}
$, where $\hat{X}, \hat{Z}$ are $\pi$-rotations about $\hat{x}, \hat{z}$.
Let the $C_2$-invariant point be the origin. Clearly, the no-go condition is unmet whenever $[\omega]_{\vec 0} = 1$, and hence it suffices to prove a no-go with $[\omega]_{\vec 0} = -1$.

\begin{figure}
\begin{center}
{\includegraphics[width=0.48 \textwidth]{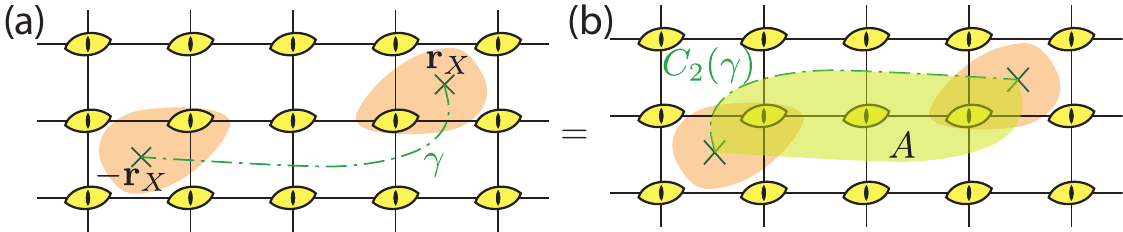}} 
\caption{{\bf Flux insertion.} (a) A $C_2$ symmetric lattice with a pair of $X$-fluxes  (crosses) inserted at $\pm \vec r_{X}$, which leads to ``defect regions'' (shaded) near the fluxes. Far away from $\pm \vec r_X$, flux insertion amounts to choosing a defect line (dash-dot) and twisting the local Hamiltonian by $X$ along the line. (b) As $X^{-1} = X$, the system retains a twisted $C_2'$ symmetry, since the transformed defect line can be brought back to the original by applying a gauge transformation on the region $A$.
\label{fig:nogo}}
\end{center}
\end{figure}

To this end, we modify the Hamiltonian by introducing a pair of $X$-fluxes at the $C_2$-related points $\pm \vec r_X$ for some arbitrarily large $|\vec r_X|$ (Fig.~\ref{fig:nogo}a).
An ``$X$-flux" is  analogous to a twist in boundary condition, and is microscopically defined as follows \cite{arXiv1410.4540}. 
We choose a line segment $\gamma$ connecting $\pm \vec r_X$, and for each local term $\hat h = \sum_{j} \hat O^j_{\rm L} \hat O^j_{\rm R}$ in the Hamiltonian intersecting $\gamma$,
where $\hat O^j_{\rm L}$ and $\hat O^j_{\rm R}$ are respectively localized to the left and right of $\gamma$,
we replace it by $\hat h' \equiv \sum_{j} \hat O^j_{\rm L}  \left( \hat X \hat O^j_{\rm R} \hat X^\dagger \right)$ to obtain  $\hat H'$.
Note that, while the flux insertion points $\pm \vec r_X$ are fixed and correspond to defects in the system, the choice of $\gamma$ is arbitrary, and one can deform $\gamma \rightarrow \gamma'$ by applying the gauge transformation $\prod_{\vec r \in A} \hat X_{\vec r}$ in the region $A$ enclosed by $\gamma' -\gamma$. Also, the orientation of $\gamma$ is immaterial as $X$ is an order-two symmetry.

Though the two fluxes are $C_2$-related, the choice of the defect line $\gamma$ naively spoils the $C_2$ symmetry.
However, the change $\gamma \rightarrow C_2(\gamma)$ can be removed by a gauge transformation (Fig.~\ref{fig:nogo}b). Consequentially, $ \hat H'$ is symmetric under a twisted-$C_2$ operation: $\hat C_2' = \left(\prod_{\vec r \in A} \hat X_{\vec r} \right) \hat C_2$, where $\partial A = \gamma - C_2(\gamma)$. 
In addition, $\hat Z \equiv \prod_{\vec r} \hat Z_{\vec r}$ remains a symmetry of $\hat H'$. Computing the commutation relation between the two symmetries, one finds
\begin{equation}\begin{split}\label{eq:C2Comm}
 \hat C_2' \hat Z \hat C_2'^{-1}\hat Z^{-1}  = \prod_{\vec r \in A}
\hat X_{\vec r}\hat Z_{\vec r} \hat X_{\vec r}^{-1}\hat Z_{\vec r}^{-1} 
=\prod_{\vec r \in A} [\omega]_{\vec r} = [\omega]_{\vec 0},
\end{split}\end{equation}
where in the last equality we used the fact that $A$ is $C_2$-symmetric, and by symmetry $[\omega]_{\vec r} = [\omega]_{-\vec r}$. Since both $\hat Z$ and $\hat C_2'$ are symmetries of $\hat H'$, they leave the ground space $\mathcal H'_{\rm GS}$ invariant. We can therefore project Eq.~\eqref{eq:C2Comm} into $\mathcal H'_{\rm GS}$, and obtain the corresponding relation $ \hat C_2'|_{\rm GS} \, \hat Z|_{\rm GS}= [\omega]_{\vec 0} \, \hat Z|_{\rm GS} \, \hat C_2'|_{\rm GS} $.

When $[\omega]_{\vec 0} = -1$, there is (at least) a two-fold degeneracy that we will now show is impossible in a sym-SRE phase, provided the degeneracy localization assumption holds. If the system was sym-SRE, degeneracy localization implied  $\hat Z|_{\rm GS} = \hat Z|_{\rm GS}^{(+)} \otimes \hat Z|_{\rm GS}^{(-)} $, where $\pm$ denotes the fluxes at $\pm \vec r_{X}$.  In addition, as $C_2$ exchanges the two fluxes, the local degeneracies satisfy $d_{+} = d_{-}$, and without loss of generality we can choose a basis in which $C_2'|_{\rm GS}$ is simply $(\hat{C}_2'|_{\rm GS} )| \alpha_+ \alpha_-\rangle=| \alpha_- \alpha_+\rangle$, where $\alpha_{\pm}$ denotes the independent degenerate states ``trapped" at $\pm \vec r_{X}$.  In this basis, the commutation relation reads
$\hat C_2'|_{\rm GS} \hat Z|_{\rm GS} \hat C_2'|_{\rm GS}^{\dagger} =\hat Z|_{\rm GS}^{(-)} \otimes \hat Z|_{\rm GS}^{(+)}=- \hat Z|_{\rm GS}^{(+)} \otimes \hat Z|_{\rm GS}^{(-)}$.
A solution to this requires $\hat Z|_{\rm GS}^{(-)} = \nu \hat Z|_{\rm GS}^{(+)}$ for some $\nu \in {\rm U}(1)$ satisfying $-\nu = \nu$, leading to a contradiction. Hence the claim.

In closing, we remark that our no-gos are circumvented if the system becomes LRE. An example is discussed in Appendix \ref{app:Toric}.

{\it Discussion and outlook.}--
In conclusion, we have conjectured that all LSM-like theorems for quantum magnets, where microscopic degrees of freedom forbid symmetric short-range entangled phases, can be understood intuitively as topological obstructions to smoothly deforming the underlying lattice into a trivial one.
We proved the conjecture in 2D for quantum magnets that are either spin-rotation invariant, or possess on-site unitary finite-Abelian symmetries. 

Our 2D arguments, in fact, cover all 80 layer groups, which are symmetries of 2D lattices embedded in 3D (Appendix \ref{app:Layer}).
They also extend to some genuinely 3D lattices   --  in particular, the three no-gos remain true, where mirror lines and $C_m$ rotation-invariant points in 2D become planes and lines in 3D.
Such extensions can have immediate implications on spin-liquid candidates. As an example, we note that both the Bieberbach and mirror no-gos are silent for the pyrochlore quantum spin ice Yb$_2$Ti$_2$O$_7$~\cite{Yb2TiO7}, but the $C_2$-rotation no-go remains active if we model the system as a spin-rotation invariant quantum magnet. Yet, we caution that spin-orbit coupling is strong in the actual material~\cite{Yb2TiO7}, and so this idealization is not immediately justified.

A closer inspection, however, reveals that these three no-gos only prove the conjecture for some but not all of the 230 3D space groups (Appendix \ref{app:3D} includes simple examples for which the current set of no-gos are insufficient.)
New techniques will be required, and we describe some partial results in Appendix \ref{app:DNA}.
We also note that it would be most useful if only time-reversal $\mathcal{T}$ was required in the no-gos, with the role of projective representation played by the Kramers degeneracy from $\mathcal{T}^2 = -1$.
However, it is not clear how to extend our flux-insertion proof to this case.
In addition, actual materials are composed of itinerant fermions carrying spin and the quantum-magnet description is often an approximation. It would be useful to know if our results extend to this more general case.
With a Mott gap, it naively seems that there should be a sharp notion of ``where" the spins of the electrons lie (at least up to the  lattice equivalence relations), but certain examples suggest this may not be the case \cite{SA, reBS}.
Finally, we note that our conjecture has interesting connection to the study of crystalline SPTs \cite{Fidkowski2015, WangLinLevin2016, ChengZaletel, HermeleChen, PhysRevX.7.011020, ThorngrenElse, ElseFuture}, which we comment briefly in Appendix \ref{app:SPT}.

\begin{acknowledgments}
We thank A. Vishwanath for discussions and collaboration on related works.
MZ is indebted to conversations with   D. Else,  M. Cheng, M. Freedman, C. Galindo-Martinez  and M. Hermele.
CMJ's research at the KITP is funded by the Gordon and Betty Moore Foundation's EPiQS Initiative through Grant GBMF4304.
\end{acknowledgments}

\bibliography{references}

\clearpage
\appendix
\section{Systematic computation of lattice homotopy using Wyckoff positions
\label{app:latticehomotopy}}
Here we will develop a systematic method to compute lattice homotopy for lattices (as summarized in Table I in the main text) symmetric under a space group $\mathcal S$ (in any spatial dimension $d$). In the following, we will always let $\eta$ be a generator (e.g. $e^{2\pi i/n}$) of $\mathcal{H}^2\left[G,U(1)\right]=\mathbb{Z}_n$.   The more general case of $\mathcal{H}^2\left[G,U(1)\right] = \prod_{i<j} \mathbb{Z}_{\gcd(n_i,n_j)}$ for the most generic finite Abelian group $G = \prod_i \mathbb{Z}_{n_i}$ will follow naturally as we discuss later.

To this end, we first introduce a notation for specifying a lattice. Suppose there is a site $\vec r$ carrying projective representation $[\omega]_{\vec r} \in \mathbb{Z}_{n}$. We will denote this site by $(\vec r ; [\omega]_{\vec r})$. For the lattice to be invariant under $\mathcal S$, for any $s \in \mathcal S$, there must be a site at $s(\vec r)$ carrying $[\omega]_{s(\vec r)} = [\omega]_{\vec r}$. We will denote the lattice generated by $(\vec r; [\omega]_{\vec r})$ by $\langle (\vec r; [\omega]_{\vec r}) \rangle \equiv \{ (s(\vec r), [\omega]_{\vec r}) : s\in \mathcal S\}$. Similarly, we will denote lattices generated by multiple sites by $\langle (\vec r; [\omega]_{\vec r}),  (\vec r'; [\omega]_{\vec r'}) \rangle \equiv 
\langle (\vec r; [\omega]_{\vec r}) \rangle + \langle (\vec r'; [\omega]_{\vec r'}) \rangle $, where the addition sign denotes lattice stacking (when sites coincide, the representations $[\omega]_{\vec r}$ and $[\omega]_{\vec r}'$ are fused by multiplication,  giving $\langle (\vec r; [\omega]_{\vec r}'[\omega]_{\vec r}) \rangle$).

From the discussion in the main text, we have seen that nontrivial lattice classes are characterized by invariants related to the total representation in special regions of space, determined by the space group $\mathcal S$.
Such special regions can be characterized through the notion of ``Wyckoff positions'', which can be viewed as a partition of space naturally arising from the action of $\mathcal S$ on $\mathbb R^d$.
A Wyckoff position is the collection of points having conjugate site-symmetry groups -- the subgroup of $\mathcal S$ that leaves the ``site'' fixed. 
To unpack this definition, consider a point $\vec x \in \mathbb R^d$ and let $\mathcal S_{\vec{x}} \equiv \{ s \in \mathcal S:s(\vec{x}) = \vec{x}\}$ denote its site-symmetry group. $\mathcal S_{\vec {x}}$ is always a subgroup of the point group $\mathcal S/T$, where $T$ denotes the translation subgroup of $\mathcal S$. 
Now let $\vec y$ be a point in the $\mathcal S$-symmetric lattice generated by $\vec x$, i.e.~$\vec y= s_0 (\vec x)$ for some $s_0 \in \mathcal S$. It is easy to check that $\mathcal S_{\vec y} = s_0 \mathcal S_{\vec x} s_0^{-1}$, i.e.~$\mathcal S_{\vec x} $ and $\mathcal S_{\vec y}$ are conjugate, and so $\vec x$ and $\vec y$ belong to the same Wyckoff position. In addition, if $\mathcal S_{\vec x}$ fixes any other points other than $\vec {x}$, say for a mirror plane in 2D, then all the lattices generated by another point in the same ``fixed set'' will also be in the same Wyckoff position. Hence, we see that Wyckoff positions can be physically interpreted as a classification of $\mathcal S$-symmetric lattices generated by a single site. In particular, note that when the corresponding site-symmetry group fixes a set instead of a single point in space, the Wyckoff position will contain free parameters parametrizing the set.

By definition, any two lattices in the same Wyckoff position can be smoothly deformed into each other without breaking symmetries.  Now if we restore the notion of projective representations carried by the sites, we see that in specifying a lattice it suffices to count the total projective representations in each of the Wyckoff position.

\begin{table}[h]
\begin{center}
\caption{The lattice homotopy classification for the 17 wallpaper groups, assuming $\mathbb{Z}_n$ projective representations. 
These results generalize readily to any $\mathcal{H}^2 \left[ G, U(1) \right] =  \prod_j \mathbb{Z}_{n_j}$.
\label{tab-Zn}}
\begin{tabular}{ll}\hline\hline
Wallpaper group No.~\cite{ITC}\hspace{10pt}\mbox{}&Lattice homotopy\\ \hline
1, 4, 5	& $\mathbb{Z}_n$\\
2, 6		& $\mathbb{Z}_n\times \left (\mathbb{Z}_{\mathrm{gcd}(n,2)}\right)^3$\\
3, 8		& $\mathbb{Z}_n\times\mathbb{Z}_{\mathrm{gcd}(n,2)}$\\
7, 9		& $\mathbb{Z}_n\times \left ( \mathbb{Z}_{\mathrm{gcd}(n,2)} \right)^2$\\
10, 11	& $\mathbb{Z}_n\times\mathbb{Z}_{\mathrm{gcd}(n,4)}\times\mathbb{Z}_{\mathrm{gcd}(n,2)}$\\
12		& $\mathbb{Z}_n\times\mathbb{Z}_{\mathrm{gcd}(n,4)}$\\
13, 14	& $\mathbb{Z}_n\times \left( \mathbb{Z}_{\mathrm{gcd}(n,3)}\right)^2$\\
15		& $\mathbb{Z}_n\times\mathbb{Z}_{\mathrm{gcd}(n,3)}$\\
16, 17	& $\mathbb{Z}_n\times\mathbb{Z}_{\mathrm{gcd}(n,3)}\times\mathbb{Z}_{\mathrm{gcd}(n,2)}$\\\hline\hline
\end{tabular}
\end{center}
\end{table}

Next, we study the relations between different Wyckoff positions.
Let $\vec x$ belong to a Wyckoff position $\mathcal W$.
If there exists another point $\vec y\in \mathbb R^d$ such that $\mathcal{S}_{\vec x}$ is a proper subgroup of $\mathcal{S}_{\vec y}$, then we say the Wyckoff position $\mathcal W$ is ``reducible.'' 
Physically, a Wyckoff position is reducible if there is a tuning parameter one can tune to bring a collection of sites to a higher symmetry point, and thereby enhancing the site-symmetry group from $\mathcal S_{\vec x}$ to $\mathcal S_{\vec y}$.
In this case, we can smoothly deform $\langle(\vec x; \eta)\rangle$ to $\langle(\vec y; \eta^p) \rangle $ by fusing $p\equiv|\mathcal{S}_{\vec y}|/|\mathcal{S}_{\vec x}|$ points at $\vec y$. 
As such, one can always trivialize the projective representation contained in a reducible Wyckoff position by ``pushing'' them to the irreducible ones.  Therefore, the lattice homotopy problem is reduced to a study of trivial elements of $\mathbb{Z}_n^{N_{\rm IWP}}$, where $N_{\rm IWP}$ denotes the number of ``irreducible'' Wyckoff positions for $\mathcal S$, and the notion of ``lattice stacking' is simply given by the group addition in $\mathbb{Z}_n^{N_{\rm IWP}}$.  We label irreducible Wyckoff positions by $\alpha=1,2,\cdots,N_{\rm IWP}$ (in the same ordering as in Supplementary Ref.~\cite{ITC}) and express a lattice as $(\eta^{q_1},\eta^{q_2},\cdots,\eta^{q_{N_{\rm IWP}}})\in \mathbb{Z}_n^{N_{\rm IWP}}$ when the total projective representations on the irreducible Wyckoff position $\mathcal{W}_\alpha$ is $\eta^{q_\alpha}$.

While any manifestly non-projective lattices will correspond to $(1,1,\dots, 1)\in \mathbb{Z}_n^{N_{\rm IWP}}$, some other entries of $\mathbb{Z}_n^{N_{\rm IWP}}$ may also be trivial under lattice homotopy, as for the example discussed in the main text, concerning the $\langle (0;\eta), (1/2;\eta^2)\rangle$ lattice 
(or equivalently, $(\eta,\eta^2)\in \mathbb{Z}_{n=3}^{N_{\rm IWP}=2}$) for the case of a 1D mirror and translation symmetric spin chain with projective representations in $\mathbb{Z}_3$. 
The trivial lattices among $\mathbb{Z}_n^{N_{\rm IWP}}$ form a group under addition, and in particular they will correspond to a subgroup $K \leq \mathbb{Z}_n^{N_{\rm IWP}}$.  To find a set of generators for $K$, we simply note the following: Let $\vec x_1$ and $\vec x_2$ belong to two different irreducible Wyckoff positions $\mathcal W_1$ and $\mathcal W_2$. Consider another point $\vec x \in \mathcal W$, and suppose  $\mathcal W$ is ``reducible'' into both $\mathcal W_1$ and $\mathcal W_2$ by setting some free parameters to special values. Physically, this describes moving some lattice sites in $\mathcal W_1$ to lattice sites in $\mathcal W_2$ via lower-symmetry points in $\mathcal W$: $\langle(\vec x_1; \eta^{p_1})\rangle\sim \langle(\vec x; \eta) \rangle\sim \langle(\vec x_2; \eta^{p_2})\rangle$ where $p_i\equiv |\mathcal{S}_{\vec x_i}|/|\mathcal{S}_{\vec x}|$. Thus one sees that $\langle (\vec x_1; \eta^{-p_1} ), (\vec x_2; \eta^{p_2}) \rangle$, which can be written as $(\eta^{-p_1},\eta^{p_2},1,\cdots,1)\in \mathbb{Z}_n^{N_{\rm IWP}}$, is trivial in the lattice homotopy sense, and therefore is a generator of $K$. All generators of $K$ can be found by checking if such reduction is possible in any pairs of irreducible Wyckoff positions. 
Once this is achieved, the lattice homotopy classification is given by the quotient $\mathbb{Z}_n^{N_{\rm WP}}/ K$. As the Wyckoff positions are well tabulated in crystallographic references like Supplementary Ref.~\cite{ITC}, the computation can be easily automated.

In Supplementary Table \ref{tab-Zn}, we tabulate the lattice homotopy classification for the 17 wallpaper groups assuming $\mathbb Z_n$ projective representations. This is a generalized version of Table I in the main text, where we restricted to $\mathbb Z_2$ projective representations. 
Note that Supplementary Table \ref{tab-Zn} generalizes readily to any $\mathcal{H}^2 \left[ G, U(1) \right] =  \prod_j \mathbb{Z}_{n_j}$.
In this case, it is easy to see that the lattice homotopy problem  factorizes over $j$, i.e., we just take a product over the classification for each $\mathbb{Z}_{n_j}$.

\section{General proof of conjecture in 2D for finite Abelian $G$
\label{app:Proof}}
Here we will prove (up to the degeneracy localization assumption) a more general version of the conjecture in 2D, where instead of considering spin-rotation invariant systems with $G={\rm SO}(3)$, we let the internal symmetry group take the form $G=\mathbb{Z}_{n_1} \times \mathbb{Z}_{n_2}$, which has $\mathcal{H}^2\left[G,U(1)\right] = \mathbb{Z}_{\gcd(n_1,n_2)}$.  In fact, for the most general finite Abelian group $G = \prod_i \mathbb{Z}_{n_i}$, the projective representations are classify by $\mathcal{H}^2\left[G,U(1)\right] = \prod_{i<j} \mathbb{Z}_{\gcd(n_i,n_j)}$. The generators of $\mathcal{H}^2\left[G,U(1)\right] $ are given by the generators of the projective representation of $\mathbb{Z}_{n_i} \times \mathbb{Z}_{n_j}$ with different choices of the pair $(i,j)$. Therefore, for the most general finite Abelian group $G$, the analysis on the no-gos follows directly from the case of $G=\mathbb{Z}_{n_1} \times \mathbb{Z}_{n_2}$, which we will focus on.

We begin by stating explicitly the version of the conjecture proven here:\\

\noindent \emph{Proposition.} Let $\mathcal S$ be a space group in 2D (one of the 17 wallpaper groups) and let the internal symmetry group be $G = \mathbb{Z}_{n_1} \times \mathbb{Z}_{n_2}$. A quantum magnet, defined on a lattice $\Lambda$ and symmetric under $\mathcal S \times G$, can be in a sym-SRE phase only if the lattice homotopy class $[\Lambda]=1$.\\

The proof takes the same structure as the simplified version presented in the main text, in which we will again assume sym-SRE phases exhibit degeneracy localization.
First, we will derive the ``Bieberbach,'' ``mirror,'' and ``rotation'' sym-SRE no-gos for $G=\mathbb{Z}_{n_1} \times \mathbb{Z}_{n_2}$. Second, we will connect the no-gos to lattice homotopy, and show at least one of the sym-SRE no-gos we derived is present if and only if $[\Lambda]\neq 1$.

Before we move on to the derivation, we first discuss the relevant structure of $G$ that leads to the no-gos.
We denote the generators of the two factors of $G=\mathbb{Z}_{n_1} \times \mathbb{Z}_{n_2} $ by $X$ and $Z$, such that $X^{n_1} = Z^{n_2}=1$.
While $X$ and $Z$ commute as group elements, their corresponding unitary operator satisfy
\begin{equation}
\hat{X}\hat{Z}\hat{X}^{-1}\hat{Z}^{-1} = \frac{\omega(X, Z)}{\omega(Z, X)} \equiv [\omega](X, Z),
\end{equation}
implying they do not commute whenever $[\omega]$ is nontrivial.  Hence, we will write $\hat X_{\vec r} \hat Z_{\vec r}  = [\omega]_{\vec r} \hat Z_{\vec r} \hat X_{\vec r}$ for local operators at site $\vec{r}$, where $[\omega]_{\vec r}\in \mathcal{H}^2\left[G,U(1)\right] = \mathbb{Z}_{n}$ and $n\equiv\gcd(n_1,n_2)$. Note that by restricting the following discussion to $\mathbb{Z}_2 \times \mathbb{Z}_2$, we recover the proof for the spin-rotation invariant magnets by viewing $G$ as the $\pi$-rotation subgroup.

\subsection{Deriving the three sym-SRE no-gos}
In the following, we will restate and then derive the three no-gos. 
First, recall that the mirror no-go and the $C_m$ rotation no-go are silent when $[\omega]_{\vec r}$ is an ``$m$-th root'' ($m=2$ for the mirror no-go), i.e., if $[\omega]_{\vec r} = \zeta^m$ for some $\zeta \in \mathcal{H}^2\left[G,U(1)\right] = \mathbb{Z}_n$. 
If we express $[\omega]_{\vec r} =\eta^q$ ($q\in \mathbb{Z}$) using a generator $\eta$ of $\mathcal{H}^2\left[G,U(1)\right]$ (e.g. $e^{2\pi i/n}$), we can rephrase this condition as 
\begin{align}
q = 0 \mod \gcd(n, m).\label{eq:nogo23}
\end{align}
To see this, note that there always exist integers $s$ and $t$ such that $s n+t m=\mathrm{gcd}(n,m)$. Hence, if Eq.~\eqref{eq:nogo23} holds, $\exists \, l$ such that
\begin{equation}
[\omega]_{\vec r}=\eta^q=\eta^{l\,\mathrm{gcd}(n,m)}=\eta^{l(s n+t m)}=\zeta^m,
\end{equation}
with $\zeta=\eta^{lt}\in\mathbb{Z}_n$.
Conversely, if $[\omega]_{\vec r} = \zeta^m$ for $\zeta=\eta^k\in\mathbb{Z}_n$, then
\begin{equation}
[\omega]_{\vec r} =\zeta^m= \eta^{km}=\eta^{l\,\mathrm{gcd}(n,m)}=\eta^q,
\end{equation}
with $l\equiv km/\mathrm{gcd}(n,m)\in\mathbb{Z}$ and $q\equiv l\,\mathrm{gcd}(n,m)$, and Eq.~\eqref{eq:nogo23} holds.

\subsubsection{Bieberbach no-go}

\noindent \emph{Bieberbach no-go:} A ``fundamental domain'' $D$ is a region which tiles the plane under the action of translation and glide symmetries. If the total projective representation in $D$ is non-trivial, $[\omega]_D \equiv \prod_{\vec r \in D} [\omega]_{\vec r} \neq 1$, then a sym-SRE phase is forbidden \cite{PNAS}.\\

The original LSM-like theorems in Supplementary Refs.~\cite{Lieb1961,Affleck1986, AffleckPRB,  YOA-PRL, Oshikawa2000,Hastings2004,Xie2011} translates the net count of $[\omega]$ in the primitive unit cell to a sym-SRE obstruction. 
In the presence of nonsymmorphic symmetries, which, like translations, move every point in space, the notion of ``unit cell'' can be generalized to a smaller region of space $D$ called the ``fundamental domain,'' which tessellates space under the action of a nonsymmorphic subgroup (more precisely, the maximal fixed-point-free subgroup) of $\mathcal S$ \cite{PNAS}.
We will now show that a sym-SRE no-go is present whenever $[\omega]_D \equiv \prod_{\vec r \in D}[\omega]_{\vec r} \neq 1$.
The full argument, which is applicable to a much more general class of models, was presented in in Supplementary Ref.~\cite{PNAS}. In the following we sketch the key ideas involved.

We begin by discussing the LSM-like theorems that utilize only the lattice translation symmetry. We take $D$ as the primitive unit cell $D_0$. Thanks to translation symmetry, the system can be consistently defined on any torus of size $L_1\times L_2 \times\dots \times L_d$. Let $T^d_V$ denote a torus of size $V\equiv\prod_{i=1}^{d} L_i$, where we have set the size of the primitive unit cell to $1$. The system defined on $T^d_V$ is symmetric under both $X$ and $Z$, which as quantum operators satisfy
\begin{equation}\begin{split}\label{eq:}
\hat X \hat Z \hat X^\dagger \hat Z^\dagger = \prod_{\vec r \in T^d_V} [\omega]_{\vec r} =[\omega]_D^{V}.
\end{split}\end{equation}
Writing $[\omega]_D=\eta^{q_D}$ with an integer $q_D$, 
we see that $\hat X$ and $\hat Z $ do not commute whenever $q_DV\neq 0 \mod n$.  Choosing $V$ to be co-prime with $n$, this implies the ground state of the system must be degenerate unless
\begin{equation}
q_D = 0 \mod n,\label{eq:nogo1}
\end{equation}
despite $V$ can be arbitrarily large. For a sym-SRE phase, however, the ground state should be gapped and unique in the thermodynamic limit, and hence we arrive at a contradiction.

The generalization to more general fixed-point-free (nonsymmorphic) symmetries proceeds in a similar manner \cite{PNAS}. The only difference is that, instead of defining the system on the torus, we define it on certain flat compact manifolds $\mathcal M$ (known as the Bieberbach manifolds). 
The Bieberbach manifolds are tied to the fixed-point-free subgroup of $\mathcal S$ that tessellates the entire space upon acting on the fundamental domain $D$. 
Focusing on a 2D system, the only nonsymmorphic symmetry is a glide mirror, which has the Klein bottle as the associated Bieberbach manifold.
One can check that, in the presence of a glide symmetry, the entire lattice can be defined by only specifying the content of half of the primitive unit cell $D_0$, so that $D$ can be identified as half of $D_0$.
The volume of the Bieberbach manifold $V_{\mathcal M}$ is generally quantized in units of $V_D$, the volume of $D$.
There a similar ground state degeneracy is exposed unless $q_D=0$ mod $n$, since one can always choose $V_{\mathcal M}$ (measured in $V_D$) to be co-prime with $n$.  Importantly, a sym-SRE phase should not be able to detect the topology of $\mathcal M$, as the Hamiltonian $\hat H_{\mathcal  M}$ defined on $\mathcal M$ is locally indistinguishable from that on the infinite space (see Supplementary Ref.~\cite{PNAS} for details). Therefore the ground state degeneracy again leads to a sym-SRE obstruction.

\subsubsection{Mirror no-go}

\noindent\emph{Mirror no-go.} Let $\ell$ be a mirror line parallel to a translation $T_\parallel$. We define the projective representation per unit length of $\ell$,  $[\omega]_\ell \equiv \prod_{\vec r \in \ell'} [\omega]_{\vec r}$, by letting the product run over a unit-length interval $\ell'$ of $\ell$ as defined by $T_{\parallel}$. If $[\omega]_\ell$  does not have a ``square-root,'' i.e., if no $\zeta\in\mathbb{Z}_n$ satisfies $[\omega]_\ell = \zeta^2$, then a sym-SRE phase is forbidden \cite{PNAS}.\\

This was again proven in Supplementary Ref.~\cite{PNAS}, but we will provide here an alternative flux-insertion argument 
as a warm-up to the rotation no-go. 
When $n$ is odd, one can always find a $\zeta \in \mathbb{Z}_n$ such that $[\omega]_\ell = \zeta^2$, as for any generator $\eta$ of $\mathbb{Z}_n$, $\eta^2$ is also a generator. Hence, it suffices to consider $G$ of the form $\mathbb{Z}_{2a} \times \mathbb{Z}_{2b}$, which gives $[\omega]_{\vec r} \in \mathbb{Z}_{n}$ with $n=2\,\mathrm{gcd}(a,b)$.
By definition $\bar a \equiv  a/{\rm gcd}(a,b)$ and $\bar b \equiv b/{\rm gcd}(a,b)$ are coprime, and without loss of generality we take $\bar a$ to be odd.

We will first argue in a 1D setting, which generalizes readily to 2D.
Consider a finite but arbitrarily large ring of spins with a mirror $m$ centered at the site $x=0$.
Now imagine inserting an $X^a$ flux through the system by twisting the boundary condition (BC) between sites $\gamma$ and $\gamma+1$. The choice of $\gamma$ is immaterial, as twisting instead at $\gamma' < \gamma$ is equivalent to applying the gauge transformation $\prod_{x\in [\gamma'+1, \gamma]} \hat X_{x}^a$.
By the degeneracy localization assumption, the flux-inserted Hamiltonian $\hat H'$ should still have a gapped, unique ground state, as we have only inserted a pure flux and there are no defects to trap degeneracies.

We will now argue a contradiction to the sym-SRE assumption when the mirror invariant is nontrivial.
The action of $\hat m$ on $\hat H'$ has two effects: first, it moves the BC twist to that between sites $-\gamma$ and $-\gamma - 1$; second, it reverses orientation and therefore the flux inserted is inverted (Supplementary Fig.~\ref{fig:fluxC4}). However, by construction $X^{-a} = X^a$, and as argued the BC twist can be brought back to $\gamma$ by a gauge transformation. This implies $\left[ \hat m', \hat H' \right] = 0$, where $\hat m'\equiv \left( \prod_{x\in [-\gamma, \gamma]} \hat X_{x}^a \right) \hat m$. As $\hat Z = \prod_{x} \hat Z_x$ remains a symmetry, we consider
\begin{equation}\begin{split}\label{eq:}
\hat m' \hat Z (\hat m')^\dagger \hat Z^\dagger = \prod_{x\in [-\gamma, \gamma]} [\omega]_{x}^a = [\omega]_0^a,
\end{split}\end{equation}
where we used the symmetry constraints to reduce $[\omega]_x^{a} [\omega]_{-x}^{a}= [\omega]_x^{2a}=1$. 
Writing $[\omega]_0 = \eta^q$ for some integer $q$, one finds $[\omega]_0^a=\eta^{n\bar{a}q/2}$. Recalling $\bar a$ is odd, we see that $\hat m' $ and $\hat Z$ commute iff $q$ is even. 
This implies $\hat H'$ has a ground state degeneracy whenever $q$ is odd. Since  $[\omega]_0 \neq \zeta^2$ for any $\zeta\in \mathbb{Z}_n$ is equivalent to $q$ being odd, the claim follows.

To generalize to the 2D case, we simply take a torus geometry with $L_{\parallel}$ unit cells in the direction parallel to the mirror line.
The net count of $[\omega]$ on the mirror line is $[\omega]_\ell^{L_{\parallel}}$, where $[\omega]_\ell$ is the count within one unit cell.
While the BC twist is now promoted from a point to a closed line, it remains invisible to a sym-SRE phase. The 1D argument goes through with the effective 1D invariant $[\omega]_{\ell}^{L_{\parallel}}$. As we are free to choose an odd $L_{\parallel}$, we conclude a sym-SRE phase is forbidden whenever the 2D mirror invariant is nontrivial.

\subsubsection{Rotation no-go}

\noindent\emph{Rotation no-go.} Let $\vec r$ be a site with rotational point-group symmetry $C_m$ and projective representation $[\omega]_{\vec r}$. If $[\omega]_{\vec r}$ does not have an ``$m$-th root,'' i.e., if no $\zeta\in\mathbb{Z}_n$ satisfies $[\omega]_{\vec r} = \zeta^m$, then a sym-SRE phase is forbidden.\\

As we have already discussed the proof for the special case of $G=\mathbb{Z}_2 \times \mathbb{Z}_2$ in the main text, here we only discuss the required generalization to $G = \mathbb{Z}_{n_1} \times \mathbb{Z}_{n_2}$.
First, we note that the condition for the no-go to be active is a directly generalization of the mirror case, where instead of an order-two symmetry (mirror) we now consider an order-$m$ one ($C_m$ rotation). 

We define the integer $a\equiv n_1/\gcd(n_1,m)$ and $b\equiv n_2/\gcd(n_2,m)$. As before, let $\eta$ denote the generator of $\mathcal{H}^2\left[G,U(1)\right]$. Then we can rewrite $[\omega]_{\vec r}$ as $[\omega]_{\vec r}= \eta^q$ for a certain integer $q$. The condition for $[\omega]_{\vec r}$ to have an ``$m$-th root'' is identical to the condition that $q$ is divisible by $\gcd(n,m)$ as discussed in Eq.~\eqref{eq:nogo23}. 

Generalizing the construction in the main text, we can insert $m$ of the $X^{a}$-fluxes at the vertices of a $C_m$-symmetric polygon $A$ centered at the origin, which requires a choice of defect line such as the one shown in Supplementary Fig.~\ref{fig:fluxC4}. Note that consistency requires $X^{a m}=1$, which is true by construction.
The flux-inserted system retains a $C_m$-symmetry  $\hat C_m' \equiv \left(\prod_{\vec{r} \in A} \hat X^{a}_{\vec{r}}  \right)\hat C_m$.
We then have 
\begin{equation}\begin{split}\label{eq:X_flux_Rot}
\hat C_m' \hat Z  (\hat {C}_m')^\dagger \hat Z^  \dagger
=  \prod_{\vec r \in A} [\omega]_{\vec r}^{a} = [\omega]_{\vec 0}^a = \eta^{aq},
\end{split}\end{equation}
where we again used the symmetry condition $[\omega]_{C_m(\vec r)} = [\omega]_{\vec r}$. If $a q = 0 ~\text{mod} ~n$, the no-go is silent in this configuration. Similarly, we can as well consider inserting $m$ of the $Z^{b}$-fluxes at the vertices of a $C_m$-symmetric polygon $A$ centered at the origin. Again, the consistent condition $Z^{b m}=1$ is automatically satisfied. This flux-inserted system also retains a $C_m$-symmetry $\hat C_m^{''} \equiv  \left(\prod_{\vec{r} \in A} \hat Z^{b}_{\vec{r}}  \right)\hat C_m$.
We then have 
\begin{equation}\begin{split}\label{eq:Z_flux_Rot}
\hat C_m^{''} \hat X (\hat C_m^{''})^\dagger \hat X^  \dagger
=  \prod_{\vec r \in A} [\omega]_{\vec r}^{b} = [\omega]_{\vec 0}^b = \eta^{bq}.
\end{split}\end{equation}
In this configuration, the no-go is silent when $b q = 0~ \text{mod} ~n$. For the no-go to be completely silent in both the configuration with $X^{a}$-fluxes and that with $Z^{b}$-fluxes, we would simultaneously need 
\begin{align}
\frac{qa}{n}=\frac{q}{\gcd(n, m)}N_1\in\mathbb{Z},\label{eq:qan}\\
\frac{qb}{n}=\frac{q}{\gcd(n, m)}N_2\in\mathbb{Z},\label{eq:qbn}
\end{align}
where $N_1\equiv\frac{n_1\gcd(n, m)}{n\gcd(n_1, m)}$ and $N_2\equiv\frac{n_2\gcd(n, m)}{n\gcd(n_2, m)}$.  Since by construction $N_1$ and $N_2$ are co-prime integers, the necessary and sufficient condition for Eqs.~\eqref{eq:qan} and \eqref{eq:qbn} is $q=0$ mod $\mathrm{gcd}(n,m)$.  [To see $N_1$ and $N_2$ are co-prime, note that $n_1/n$ ($n_2/n$) is divisible by $N_1$ ($N_2$) and that $n_1/n$ and $n_2/n$ are co-prime.]  As discussed above, this is equivalent with the absence of ``$m$-th root'' for $[\omega]_{\vec r}$.

\begin{figure}
{\includegraphics[width=0.35\textwidth]{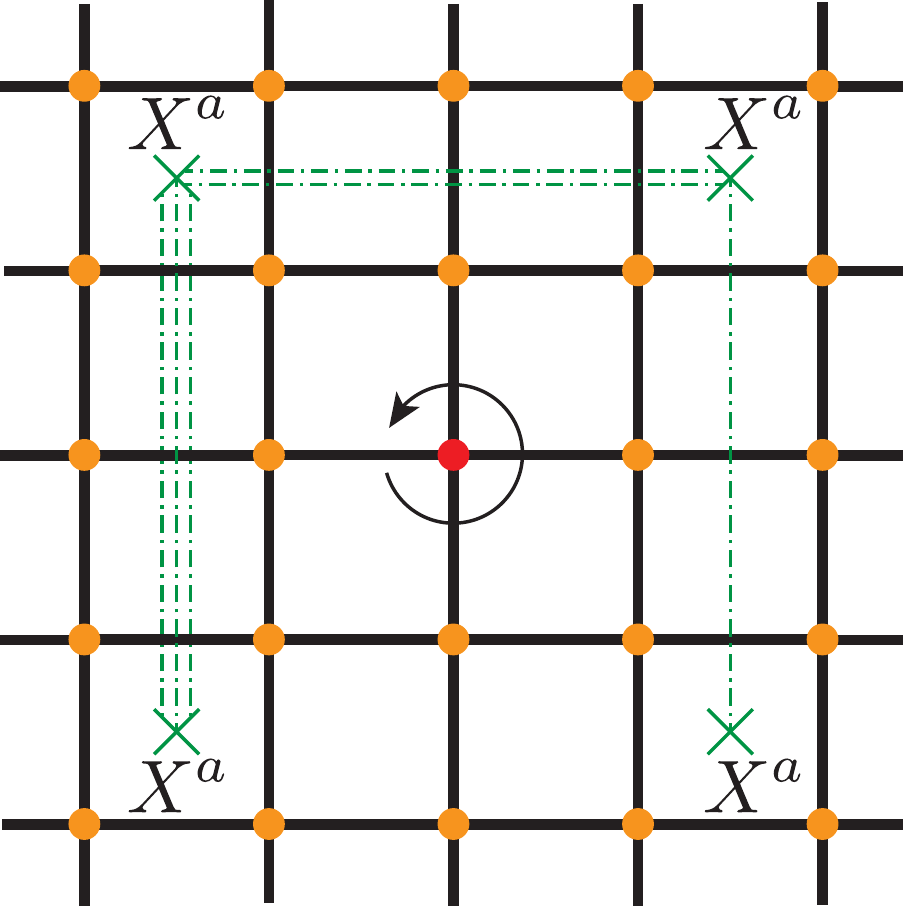}}
\caption{A $C_4$ symmetric lattice with all the spins living on the sites (orange and red). The rotation center is specified by the red site. $X^a$ fluxes are inserted at 4 locations related to each other by the $C_4$ rotation. The dotted lines are the branch cut associated to the fluxes. The symmetry action associated with each dotted line is given by $X^a$ (e.g.~that for the tripped dotted line is $X^{3a}$). The consistency condition requires $X^{4a} = 1$.
\label{fig:fluxC4}
} 
\end{figure}

Without loss of generality, we can just study the case when $[\omega]_{\vec 0}^a \neq 1$. As before, we must show this degeneracy cannot be localized. 
This follows from a direct generalization of the argument in the main text. Suppose on the contrary that the degeneracies are localized. Then we can project the symmetry operators $\hat C_m'$ and $\hat Z$ into the ground space, and with a proper choice of basis obtains
\begin{equation}\begin{split}\label{eq:}
\hat Z|_{\rm GS} =& \hat Z|_{\rm GS}^{(1)} \otimes \hat Z|_{\rm GS}^{(2)}  \otimes \dots \otimes \hat Z|_{\rm GS}^{(m)};\\
\hat C_m'|_{\rm GS} =&\hat P_{(1,2,\dots,m)},\\
\end{split}\end{equation}
where $\hat P_{(1,2,\dots,m)}$ cyclically permutes the $m$ defect regions. Computing $\hat C_m'|_{\rm GS} \hat Z|_{\rm GS}\hat C_m'|_{\rm GS}^\dagger$ in two ways, we demand
\begin{equation}\begin{split}\label{eq:CmDegLoc}
& [\omega]_0^{a} \hat Z|_{\rm GS}^{(1)} \otimes \hat Z|_{\rm GS}^{(2)}  \otimes \dots \otimes \hat Z|_{\rm GS}^{(m)}\\
= & \hat Z|_{\rm GS}^{(m)} \otimes \hat Z|_{\rm GS}^{(1)}  \otimes \dots \otimes \hat Z|_{\rm GS}^{(m-1)},
\end{split}\end{equation}
which implies $ \hat Z|_{\rm GS}^{(1)}\propto  \hat Z|_{\rm GS}^{(2)}\propto \dots  \propto  \hat Z|_{\rm GS}^{(m)}$, where $\propto$ here means that the two sides are identical up to a ${\rm U}(1)$ phase. We can therefore write $\hat Z|_{\rm GS}^{(j)} = \nu_j \hat Z|_{\rm GS}^{(1)}$, where $\nu_j\in {\rm U}(1)$ for $j=2,\dots, m$. Substituting this into the two sides of Eq.~\eqref{eq:CmDegLoc}, one finds
\begin{equation}\begin{split}\label{eq:}
[\omega]_0^{a} \nu_2 \nu_3\dots \nu_m = \nu_2 \nu_3\dots \nu_m ,
\end{split}\end{equation}
an impossibility as $[\omega]_0^a\neq 1$. Therefore, we proved that for $G=\mathbb{Z}_{n_1} \times \mathbb{Z}_{n_2}$, the absence of an ``$m$-th root'' for the projective representation $[\omega]_{\vec r}$ is the necessary and sufficient condition for the existence of a no-go theorem, via flux inserted configurations, to a sym-SRE phase.

\subsection{No-gos for the most generic finite Abelian group $G$}
For the most general finite Abelian group $G = \prod_i \mathbb{Z}_{n_i}$, the analysis on the no-gos to sym-SRE follows directly from the case with $G=\mathbb{Z}_{n_i} \times \mathbb{Z}_{n_j}$ for each pair of $(i,j)$. Hence, the absence of an ``$m$-th root'' for the projective representation $[\omega]_{\vec r}$ is again the necessary and sufficient condition for the existence of a no-go theorem to a sym-SRE phase.

\subsection{The technical obstacle for a general $G$}
In this work, we only consider the case of a finite Abelian group $G$. For a generic group $G$, we can only rely on our analysis of no-gos for any finite Abelian subgroup of $G$. An active no-go for any finite Abelian subgroup definitely implies a no-go for the full symmetry group $G$. However, we are unable to prove that the silence of no-gos for all finite Abelian subgroups implies the silence of no-gos for the full symmetry group $G$, which presents an obstacle for extending our general result to a general group $G$.

The essence of the proof for Abelian $G$ was to examine each projective commutation relation $\hat{X} \hat{Z} \hat{X}^{-1} \hat{Z}^{-1} = [\omega](X,Z)$, which, for Abelian $G$, completely characterizes the projective rep.
The case with a general group $G$ is much more complicated, since, if the pair does not commute, the commutator is not a gauge-invariant quantity.

For a general projective representation $[\omega] \in \mathcal{H}^2\left[G,U(1)\right]$, we can still study the projective commutation relations for every  commuting pairs of group elements $g,h\in G$, i.e. $gh=hg$:
\begin{align}
\hat{g} \hat{h} = [\omega](g,h) \, \hat{h} \hat{g}, 
\end{align}
where $\hat{g}$ and $\hat{h}$ are the operators in the projective representation $[\omega]$ associated to the group elements $\hat{g}$ and $\hat{h}$, and $[\omega](g,h) \in {\rm U}(1)$ is a phase factor. This relation is a direct generalization of the Abelian case. Supplementary Ref.~\cite{Miller1952} points out that $[\omega](g,h)$ as a ${\rm U}(1)$-valued function on all the commuting pairs of group elements $(g,h)$ {\it fully} specifies the projective representation $[\omega] \in \mathcal{H}^2\left[G,U(1)\right]$ for any finite group $G$. 

For a general group $G$, each commuting pair of group elements $g,h\in G$ allow us to define an Abelian subgroup $\mathbb{A}_{g,h} \subseteq G$ generated by $g$ and $h$, and one can view $[\omega](g,h)$ as a projective representation in $\mathcal{H}^2\left[\mathbb{A}_{g,h},U(1)\right]$. The analysis for the rotation no-go theorem of sym-SRE with Abelian symmetry group $\mathbb{Z}_{n_1} \times \mathbb{Z}_{n_2}$ can be directly applied to the subgroup $\mathbb{A}_{g,h}$. This is because a sym-SRE with global symmetry $G$ can also be viewed as a sym-SRE with global symmetry $\mathbb{A}_{g,h} \subseteq G$. If $[\omega]$ has an $m$-th root in $\mathcal{H}^2\left[G,U(1)\right]$, $[\omega](g,h)$ naturally has an $m$-th root in $\mathcal{H}^2\left[\mathbb{A}_{g,h},U(1)\right]$ for each commuting pairs $(g,h)$. In this case, all the rotation no-gos, which are derived from the configurations with fluxes inserted, for all the Abelian subgroup symmetry $\mathbb{A}_{g,h}$ are silent. On the other hand, if, for certain commuting pairs $(g,h)$, $[\omega](g,h)$ does not admit an $m$-th root in $\mathcal{H}^2\left[\mathbb{A}_{g,h},U(1)\right]$, a sym-SRE phase under $\mathbb{A}_{g,h}$ is forbidden, which then implies a no-go for sym-SRE phase with global symmetry $G$.

However, we cannot prove that, if $[\omega]$ does not have an $m$-th root in $\mathcal{H}^2\left[G,U(1)\right]$, there must exist a pair $(g,h)$ such that $[\omega](g,h)$ does not have an $m$-th root in $\mathcal{H}^2\left[\mathbb{A}_{g,h},U(1)\right]$, or equivalently there must exist an Abelian subgroup symmetry $\mathbb{A}_{g,h}$ that provides a no-go.
This is a purely mathematical question (which we would appreciate help with :).
Nevertheless, the fact that the projective representation $[\omega] \in \mathcal{H}^2\left[G,U(1)\right]$ can be completely specified by the $[\omega](g,h)$ with commuting pair $(g,h)$ for all finite group $G$, combined with the fact that when $[\omega]= \zeta^m$ for some $\zeta \in \mathcal{H}^2\left[G,U(1)\right]$ the rotation no-go is silent, is suggestive that the absence of the $m$-th root of $[\omega]$ is tied to the existence of a rotation no-go.

\subsection{Relating sym-SRE no-gos to lattice homotopy
\label{app:LH=NG}}
Thus far, we have defined the lattice homotopy equivalence relation $\sim_{\rm LH}$, shown how to compute the resulting classes $[\Lambda]$, and independently derived a set of sym-SRE no-gos. We now show that a no-go is present if an only if $[\Lambda] \neq 1$.
To this end, we note that the no-gos themselves suggest an equivalence relation: Given a set of sym-SRE no-gos, we say two lattices $\Lambda_1 \sim_{\rm NG} \Lambda_{2}$ if the lattice $\Lambda_1 - \Lambda_2$ is no-go-free, where the minus sign ``$-$" denotes the operation of inverting all the projective representations in $\Lambda_2$ and then stacking it with $\Lambda_1$. Note that $\sim_{\rm NG} $ can be defined for \emph{any} set of sym-SRE no-gos which inherit a natural group structure from lattice stacking. In general, when fewer no-gos are employed a coarser classification is obtained.  
Here, we will clarify the relation between $\sim_{\rm LH}$ and $\sim_{\rm NG}$ for the case $\mathcal{H}^2\left[G,U(1)\right]=\mathbb{Z}_n$ and the three no-gos we derived.

We will again work in terms of irreducible Wyckoff positions. We showed in Sec.\ \ref{app:latticehomotopy} of this supplementary note that it is sufficient to consider only lattices defined on irreducible Wyckoff positions $\mathbb{Z}_n^{N_{\rm IWP}}$ to compute the lattice homotopy.  In particular, the lattice homotopy classification can be achieved by finding the subgroup $K_{\rm LH}$ of $\mathbb{Z}_n^{N_{\rm IWP}}$ that includes all trivial lattices under lattice homotopy: 
\begin{equation}
K_{\rm LH}=\{\Lambda\in \mathbb{Z}_n^{N_{\rm IWP}}\,|\,\Lambda\sim_{\rm LH}(1,1,\cdots,1)\}.
\end{equation}
By the same token, the lattice classification under $\sim_{\rm NG}$ can be computed by finding the subgroup $K_{\rm NG}$ that corresponds to lattices without any of the three sym-SRE no-gos we have derived:
\begin{equation}
K_{\rm NG}=\{\Lambda\in \mathbb{Z}_n^{N_{\rm IWP}}\,|\,\Lambda\sim_{\rm NG}(1,1,\cdots,1)\}.
\end{equation}
Note that 
\begin{equation}
K_{\rm LH} \subseteq K_{\rm NG} \label{eq:Krel1}
\end{equation}
automatically follows in general regardless of the dimensionality of the space or the set of sym-SRE no-gos, as long as one assumes the stability of no-gos against smooth deformations of a lattice.  Namely, if $\Lambda_1$ is smoothly deformable into $\Lambda_2$ while respecting all symmetries, any no-gos applicable to $\Lambda_1$ should also be applicable to $\Lambda_2$ and vice versa.   Thus, $\Lambda_1\sim_{\rm LH}\Lambda_2$ indicates $\Lambda_1\sim_{\rm NG}\Lambda_2$.  Therefore, to establish $\sim_{\rm LH}\,=\,\sim_{\rm NG}$ it remains to show $K_{\rm LH} = K_{\rm NG}$ based on the three no-gos derived above. Note that these no-gos are not enough to show $K_{\rm LH} = K_{\rm NG}$ for all space groups in 3D (Sec.\ \ref{app:3D}), and one needs to add more to achieve the equality.

\subsubsection{$K_{\rm LH}$}
Let us determine $K_{\rm LH}$ by deriving the conditions on when a lattice $\Lambda=(\eta^{q_1},\eta^{q_2},\cdots,\eta^{q_{N_{\rm IWP}}})\in \mathbb{Z}_n^{N_{\rm IWP}}$ is trivial in the lattice homotopy sense.  

As discussed before, if there exists a Wyckoff position $\mathcal W$ reducible into both $\mathcal{W}_\alpha$ and $\mathcal{W}_\beta$, we have $\langle(\vec x_\alpha; \eta^{p_\alpha})\rangle\sim_{\rm LH}\langle(\vec x_\beta; \eta^{p_\beta})\rangle$ where $p_\alpha|\mathcal{W}_\alpha|=p_\beta|\mathcal{W}_\beta|$.  Here, $|\mathcal{W}_\alpha|$ denotes the number of sites per unit cell of a lattice belonging to the Wyckoff position $\mathcal{W}_\alpha$.  It is easy to show that $|\mathcal{S}_{\vec x_\alpha}|\times|\mathcal{W}_\alpha|$ always coincides with the order of the point group $|\mathcal{S}/T|$.  Moreover, if there is also a relation $\langle(\vec x_\beta; \eta^{p_\beta})\rangle\sim_{\rm LH}\langle(\vec x_\gamma; \eta^{p_\gamma})\rangle$, then one can combine the two relations to get  $\langle(\vec x_\alpha; \eta^{p_\alpha})\rangle\sim_{\rm LH}\langle(\vec x_\gamma; \eta^{p_\gamma})\rangle$.  By repeating this process, one can move sites in Wyckoff position $\mathcal{W}_\alpha$ to $\mathcal{W}_\beta$, and subsequently to $\mathcal{W}_\gamma$ and so forth.  

Now let us fix $\alpha>1$ and find the minimum of $p_{\alpha}$ in these relations. We denote the minimum by $n_\alpha$. Here, $n_\alpha$ can be either $2$, $3$, $4$, or $6$.  By studying Supplementary Ref.~\cite{ITC}, we find the following nice properties in 2D: (i) The minimum is achieved when the destination is the most-symmetric Wyckoff position $\mathcal{W}_1$: $\langle(\vec x_\alpha; \eta^{n_\alpha})\rangle\sim_{\rm LH}\langle(\vec x_1; \eta^{n_{\alpha}|\mathcal{W}_\alpha|/|\mathcal{W}_1|})\rangle$ or
\begin{equation}
 (\eta^{-n_{\alpha}|\mathcal{W}_\alpha|/|\mathcal{W}_1|},1,\cdots,1,\eta^{n_{\alpha}},1,\cdots,1)\in K_{\rm LH}.\label{ap:alphato1}
\end{equation}
(ii) The set of Eq.~\eqref{ap:alphato1} for each $\alpha>1$ generates $K_{\rm LH}$. Namely, there is no element of $K_{\rm LH}$ that cannot be obtained by combining the relations of the form of Eq.~\eqref{ap:alphato1}.  

Armed with these facts in 2D, we now come back to the question of when a lattice $\Lambda=(\eta^{q_1},\eta^{q_2},\cdots,\eta^{q_{N_{\rm IWP}}})\in \mathbb{Z}_n^{N_{\rm IWP}}$ is trivial in the lattice homotopy classification. If $q_\alpha$ ($\alpha>1$) is an integer multiple of $\mathrm{gcd}(n,n_\alpha)$, one can completely trivialize the projective representations at $\mathcal W_\alpha$ by moving them to $\mathcal W_1$.  After moving all projective representations to $\mathcal W_1$, the lattice becomes $\langle(\vec x_1; \eta^{\sum_{\alpha=1}^{N_{\rm IWP}}q_\alpha|\mathcal{W}_\alpha|/|\mathcal{W}_1|})\rangle$, which is trivial only when the exponent is an integer multiple of $n$.  Thus a lattice is trivial (i.e., an element of $K_{\rm LH}$) if and only if the following two conditions are simultaneously met
\begin{description}
\item[C1] $q_\alpha=0$ mod $\mathrm{gcd}(n,n_\alpha)$ for $\alpha=2,3,\dots,N_{\rm IWP}$.
\item[C2] $\sum_{\alpha=1}^{N_{\rm IWP}}q_\alpha|\mathcal{W}_\alpha|/|\mathcal{W}_1|=0$ mod $n$.
\end{description}
The lattice homotopy classification $\mathbb{Z}_n^{N_{\rm IWP}}/ K_{\rm LH}$ is thus given by
\begin{equation}
\mathbb{Z}_n\times\mathbb{Z}_{\mathrm{gcd}(n, n_2)}\times\mathbb{Z}_{\mathrm{gcd}(n, n_3)}\times\cdots\times\mathbb{Z}_{\mathrm{gcd}(n, n_{N_{\rm IWP}})},
\end{equation}
which leads to Supplementary Table~\ref{tab-Zn}. Note again that this expression is not generally valid in higher dimensions, although the lattice homotopy classification will always be a finite abelian group (assuming $\mathbb Z_n$ projective classes).

\subsubsection{$K_{\rm NG}$}
Let us move on to $K_{\rm NG}$. We ask when a lattice $\Lambda=(\eta^{q_1},\eta^{q_2},\cdots,\eta^{q_{N_{\rm IWP}}})\in \mathbb{Z}_n^{N_{\rm IWP}}$ is free from the Bieberbach, mirror and rotation sym-SRE no-gos.  

Let us start with the Bieberbach no-go. Recall that we have a projective representation $\eta^{q_\alpha}$ on each site of IWP $\mathcal{W}_\alpha$ and there are $|\mathcal{W}_\alpha|$ such sites in a primitive unit cell.  Therefore, the total projective representations in a primitive unit cell $D_0$ is given by $\prod_{\vec r \in D_0} [\omega]_{\vec r}=\eta^{\sum_{\alpha=1}^{N_{\rm IWP}}q_\alpha|\mathcal{W}_\alpha|}$.
However, in the presence of the glide symmetry the true fundamental domain $D$ is reduced by the factor of $1/|\mathcal{W}_1|$ from $D_0$ (note that this may not hold in 3D \cite{PNAS}). Hence, the total projective representation in $D$, $[\omega]_D\equiv\prod_{\vec r \in D} [\omega]_{\vec r}=\eta^{q_D}$, satisfies $q_D=\sum_{\alpha=1}^{N_{\rm IWP}}q_\alpha|\mathcal{W}_\alpha|/|\mathcal{W}_1|$.  The ``Bieberbach'' no-go is applicable whenever $q_D\neq 0$ mod $n$, see Eq.~\eqref{eq:nogo1}.

Next, let us discuss the mirror and rotation no-gos. We will introduce another integer $m_\alpha=2$, $3$, $4$, or $6$ for each IWP, and we say a mirror plane is ``effective" when the mirror plane includes only one IWP and contains an odd number of sites. 
Now we make three observations:
(i) When an IWP is on a $C_m$ rotation axis but is not on any effective mirror plane, $m_\alpha=m$; (ii) When an IWP is on an effective mirror plane but is not symmetric under any rotation, $m_\alpha=2$; (iii) When an IWP is on a $C_m$ rotation axis and is also on at least one effective mirror plane, $m_\alpha=\text{lcm}(m,2)$.  We list $m_\alpha$ and $|\mathcal{W}_\alpha|$ for each IWP for all 17 wallpaper groups in Supplementary Table~\ref{tab2}.  The mirror and rotation no-gos are effective when $q_\alpha\neq0$ mod $\mathrm{gcd}(n, m_\alpha)$ for some $\alpha$, see  Eq.~\eqref{eq:nogo23}.  Even if $q_\alpha=0$ mod $\mathrm{gcd}(n, m_\alpha)$ for every $\alpha$, it might still be possible that an ineffective mirror leads to no-go.  However, as we will see shortly we do not have to worry about this possibility.

To summarize, the \emph{necessary} conditions to go around the three sym-SRE no-gos are
\begin{description}
\item[C1'] $q_\alpha=0$ mod $\mathrm{gcd}(n, m_\alpha)$ for $\alpha=2,3,\dots,N_{\rm IWP}$.
\item[C2] $\sum_{\alpha=1}^{N_{\rm IWP}}q_\alpha|\mathcal{W}_\alpha|/|\mathcal{W}_1|=0$ mod $n$.
\end{description}
Here we intentionally dropped the case $\alpha=1$ from {\bf C1'}.  Let $K_{\rm NG}'$ be the set of lattices satisfying {\bf C1'}, {\bf C2}. These conditions may not be sufficient because we dropped the case $\alpha=1$ and also because we do not take into account ineffective mirrors.  Hence, $ K_{\rm NG}'$ gives only an ``upper bound'' of $ K_{\rm NG}$:
\begin{equation}
K_{\rm NG}\subseteq K_{\rm NG}' \label{eq:Krel2}
\end{equation}

\subsubsection{Relation between $K_{\rm LH}$ and $K_{\rm NG}$}
Staring at the above expressions, one sees that conditions {\bf C1}, {\bf C2} and {\bf C1'}, {\bf C2} are almost the same.  We see that {\bf C1} and {\bf C1'} differ only by the fact that $n_\alpha$ in {\bf C1} is replaced by $m_\alpha$ in {\bf C1'}.  By studying Supplementary Ref.~\cite{ITC}, we found that, in fact, $n_\alpha=m_\alpha$ for all wallpaper groups and IWPs {\em except} for the wallpaper group No.~12.  Therefore, for all wallpaper groups but one, we have
\begin{equation}
K_{\rm LH}=K_{\rm NG}' \label{eq:Krel3}.
\end{equation}
Combining Eqs.~\eqref{eq:Krel1}, \eqref{eq:Krel2}, and \eqref{eq:Krel3}, we establish $K_{\rm LH}=K_{\rm NG}$ ($=K_{\rm NG}'$) and hence $\sim_{\rm LH}\,=\,\sim_{\rm NG}$.

For the wallpaper group No.~12, one needs (i) $q_1=0$ mod $\mathrm{gcd}(n, 4)$, (ii) $q_2=0$ mod $\mathrm{gcd}(n, 2)$, and (iii) $q_1+q_2=0$ mod $n$ to go around the three no-gos.  It is easy to check that conditions (i)--(iii) recover {\bf C1} and {\bf C2}, and hence $K_{\rm LH}=K_{\rm NG}$ by the same logic.

\begin{table}
\begin{center}
\caption{IWP data for each of the 17 wallpaper groups.  \label{tab2}}
\begin{tabular}{ccccc}\hline\hline
WG No.&$(m_1,|\mathcal{W}_1|)$&$(m_2,|\mathcal{W}_2|)$&$(m_3,|\mathcal{W}_3|)$&$(m_4,|\mathcal{W}_4|)$\\ \hline
1		&$(1,1)$	&--		&--		&--		\\
2		&$(2,1)$	&$(2,1)$	&$(2,1)$	&$(2,1)$	\\
3		&$(2,1)$	&$(2,1)$	&--		&--		\\
4		&$(1,2)$	&--		&--		&--		\\
5		&$(2,1)$	&--		&--		&--		\\
6		&$(2,1)$	&$(2,1)$	&$(2,1)$	&$(2,1)$	\\
7		&$(2,2)$	&$(2,2)$	&$(2,2)$ 	&--		\\
8		&$(2,2)$	&$(2,2)$	&--		&--		\\
9		&$(2,1)$	&$(2,1)$	&$(2,2)$	&--		\\
10		&$(4,1)$	&$(4,1)$	&$(2,2)$	&--		\\
11		&$(4,1)$	&$(4,1)$	&$(2,2)$	&--		\\
12		&$(4,2)$	&$(2,2)$	&--		&--		\\
13		&$(3,1)$	&$(3,1)$	&$(3,1)$	&--		\\
14		&$(3,1)$	&$(3,1)$	&$(3,1)$	&--		\\
15		&$(6,1)$	&$(3,2)$	&--		&--		\\
16		&$(6,1)$	&$(3,2)$	&$(2,3)$	&--		\\
17		&$(6,1)$	&$(3,2)$	&$(2,3)$	&--		\\\hline\hline
\end{tabular}
\end{center}
\end{table}

\section{Discussions on degeneracy localization
\label{app:DegLoc}}
Here we discuss the details of the notion of ``degeneracy localization," and demonstrate how it enables us to pass the locality structure of the full Hilbert space to the ground-state subspace. 

\subsection{Entanglement structure of $\mathcal H_{\rm GS}'$}
For concreteness, let $\hat H$ be the original, defect-free Hamiltonian defined on a boundaryless geometry.
Now consider a Hamiltonian $\hat H'$ identical to $\hat H$ except for the presence of defects at two well-separated finite regions, $A$ and $B$ (see the main text for the definition of ``defect region'').
For simplicity, we consider the case of two defect regions in the following, although the arguments apply directly to the case of multiple regions.

Pictorially, a sym-SRE phase is incapable of detecting the presence of the defects at length scales much larger than the correlation length $\xi$, and therefore if $\hat H'$ exhibits ground-state degeneracy due to the presence of defects, these degeneracies cannot be ``shared," i.e.~they should be independently trapped within each defect region. This is the idea behind the ``degeneracy localization'' assumption, which we make precise below.

Let $\mathcal H_{\rm GS}'$ be the ground-state subspace of $\hat H'$, which due to the presence of defects is generally \emph{not} $1$-dimensional. We say the system exhibits \emph{degeneracy localization} if 
\begin{enumerate}[(i)]
\item There exists an orthonormal basis 
\begin{equation}\begin{split}\label{eq:}
\mathcal H_{\rm GS}' = \text{span} \{ |\alpha, \beta \rangle ~:~ \alpha=1,\dots, d_{A} ; \beta=1,\dots, d_B\rangle;
\end{split}\end{equation}
and
\item For any normalized $d_A$-component vector $\psi$, we define a state in $\mathcal H_{\rm GS}'$
\begin{equation}\begin{split}\label{eq:app_psibeta}
|\psi,\beta \rangle \equiv \sum_{\alpha=1}^{d_A} | \alpha, \beta\rangle \psi_{\alpha}.
\end{split}\end{equation}
Then for any $\psi$ and a ``reference'' basis state labeled by $\alpha_0$, there exists a local unitary operator $ \hat U^A_{(\psi, \alpha_0)}$, acting nontrivially only in $A$, such that 
\begin{equation}\begin{split}\label{eq:app_unitary}
|\psi,\beta\rangle =  \hat U^A_{(\psi, \alpha_0)} |  \alpha_0, \beta\rangle.
\end{split}\end{equation}
Note that $\alpha_0$ on the right-hand side is \emph{not} summed over. We also assume the corresponding conditions for $B$.
\end{enumerate}
In words, (i) formalizes the notion that the ground-state degeneracy of $\hat H'$ arises from that trapped at $A$ and $B$, and in particular $\dim \mathcal H_{\rm GS}' = d_A d_B$;  (ii) formalizes the idea that the degeneracies in $A$ and $B$ are independent, in the sense that one can freely rotate between the individual degeneracies using local unitary operators. Note that in general the unitary operators $\hat U^{A}_{(\psi,\alpha_0)}$ are only exponentially localized in $A$ with the localization length $\xi$. In principle, we can enlarge the size of $A$ and $B$ to maximize the ``localization,'' but the sizes of the defect regions are ultimately limited by $r$, the separation between $A$ and $B$. Hence, the discussion below carries an exponentially small correction $\mathcal O(e^{- r / \xi})$, which we will not address carefully.

Intuitively, the notion of ``degeneracy localization'' suggests  $\mathcal H_{\rm GS}' \simeq \mathcal H_{A} \otimes \mathcal H_B$, where $\dim \mathcal H_{A,B} = d_{A,B}$. While this serves as a useful mental picture, this factorization cannot be (immediately) taken literally, since $\mathcal H_{\rm GS}'$, being a subspace of the full Hilbert space, has more structure than this simple formula entails.
(More accurately, the only subtlety lies in the physical interpretation of a ``cut'' separating $\mathcal H_A$ and $\mathcal H_B$, since up to isomorphism there is only one Hilbert space of the finite dimension $d_A d_B$. 
The concern here is that the tensor product $\otimes $ does not automatically coincide with the physical one defined using the underlying lattice.)

To see this more explicitly, consider an entanglement cut with respect to an arbitrarily large region $\mathcal R$, with linear size $l_{\mathcal R} \sim r \gg \xi$, containing $A$ but not $B$.
The Schmidt decomposition for a particular basis state $|\alpha_0,\beta_0\rangle \in \mathcal H_{\rm GS}'$ reads 
\begin{equation}\begin{split}\label{eq:app_Schmidt}
|\alpha_0, \beta_0 \rangle = \sum_{i} | \alpha_0; i\rangle | \beta_0 ; i \rangle s_i ,
\end{split}\end{equation}
where  $\langle \alpha_0; i' | \alpha_0; i\rangle = \langle \beta_0; i' | \beta_0, i\rangle = \delta_{i',i}$.

By the degeneracy localization assumption, any other states $|\psi, \varphi \rangle$ of $\mathcal H_{\rm GS}'$ will have the same entanglement spectrum $ \{ s_i\} $. This can be seen from the reduced density matrix
\begin{equation}\begin{split}\label{eq:rhoU}
\left( \hat \rho_{\psi, \varphi}\right)_{\mathcal R} \equiv& \text{Tr}_{\bar{  \mathcal R}} \left( | \psi, \varphi \rangle \langle \psi, \varphi | \right)\\
=& \text{Tr}_{\bar{  \mathcal R}} \left( \hat U^{B}_{(\varphi,\beta_0)}  | \psi, \beta_0  \rangle \langle \psi , \beta_0 |   \hat U^{B \dagger}_{(\varphi,\beta_0)} \right)\\
=& \text{Tr}_{\bar{  \mathcal R}} \left( \hat U^A_{(\psi, \alpha_0)}  | \alpha_0, \beta_0  \rangle \langle \alpha_0 , \beta_0 |   \hat U^{A\dagger}_{(\psi, \alpha_0)} \right)\\
=&  \hat U^A_{(\psi, \alpha_0)} \, \text{Tr}_{\bar{  \mathcal R}} \left(  | \alpha_0, \beta_0  \rangle \langle \alpha_0 , \beta_0 |  \right) \, \hat U^{A\dagger}_{(\psi, \alpha_0)} \\
=& \hat U^A_{(\psi, \alpha_0)} \, 
\left( \hat \rho_{\alpha_0, \beta_0}\right)_{\mathcal R} 
\, \hat U^{A\dagger}_{(\psi, \alpha_0)},
\end{split}\end{equation}
where in going to the third line we have used the cyclic property of the trace, and in going to the fourth we used the assumption that $\hat U^A_{(\psi, \alpha_0)}|_{\bar{  \mathcal R}} = \hat 1$. Since $\left( \hat \rho_{\psi,\varphi}\right)_{\mathcal R} $ and $\left( \hat \rho_{\alpha_0,\beta_0}\right)_{\mathcal R} $ are unitarily-related, they have the same spectrum, i.e.~all the states $| \psi, \varphi \rangle$ share the same entanglement spectrum $\{ s_i\}$.
This is expected, since for these states the entanglement spectrum w.r.t.~$\mathcal R$ should be indistinguishable from the sym-SRE ground state of the original Hamiltonian $\hat H$, which we denote by $|0\rangle$, up to an exponential accuracy $\mathcal O(e^{- l_{\mathcal R} / \xi})$.

Now consider a state $| \psi,\beta_0\rangle$. By  Eqs.~\eqref{eq:app_psibeta} and \eqref{eq:app_Schmidt}, we get
\begin{equation}\label{eq:app_theotherhand}
|\psi,\beta_0 \rangle=\sum_{i} \left( \sum_{\alpha}  | \alpha; i \rangle\psi_\alpha \right) |\beta_0; i \rangle s_i .
\end{equation}
From the discussion in Eq.~\eqref{eq:rhoU}, we see that the Schmidt states of $|\psi,\beta_0\rangle$ in $\bar {\mathcal R}$ coincide with that of $| \alpha_0,\beta_0\rangle$. By the orthonormality of the Schmidt states, we then see that the Schmidt states in $R$ are given by $|\psi;i\rangle = \sum_{\alpha}  | \alpha; i \rangle\psi_\alpha $. On the other hand, Eq.~\eqref{eq:rhoU} shows that the reduced density matrices $ (\hat \rho_{\psi,\beta_0})_{\mathcal R}$ and $  (\hat \rho_{\alpha_0,\beta_0})_{\mathcal R}$, and hence the Schmidt states, are related by the unitary $\hat U^A_{(\psi,\alpha_0)}$. This therefore implies
\begin{equation}
|\psi;i\rangle = \hat U^{A}_{(\psi,\alpha_0)} | \alpha_0 ; i \rangle=\sum_{\alpha}  | \alpha; i \rangle \psi_\alpha,
\end{equation}
which basically allows us to ``carry over'' the similar relation from the original many-body state to the Schmidt states.

Being Schmidt states, $  |\psi; i\rangle  $ for different $i$'s are orthonormal.
This has an important consequence, which sharpens our notion of ``independent degeneracies.'' Consider
\begin{equation}\begin{split}\label{eq:app_Mij}
\delta_{i,j} = \langle \psi; i | \psi; j \rangle
=\sum_{\alpha,\alpha'} \psi^*_{\alpha'} \langle \alpha' ; i | \alpha; j \rangle \psi_\alpha 
= \psi^\dagger M^{i,j} \psi,
\end{split}\end{equation}
where we defined  $(M^{i,j})_{\alpha',\alpha} \equiv \langle \alpha' ; i | \alpha; j \rangle$.  
Now we extract the Hermitian parts by defining
\begin{equation}\begin{split}\label{eq:}
M^{(i,j)}\equiv& \frac{1}{2} \left( M^{i,j} + M^{j,i} \right);\\
M^{[i,j]}\equiv& \frac{i}{2} \left( M^{i,j} - M^{j,i} \right).
\end{split}\end{equation}
Observe that Eq.~\eqref{eq:app_Mij} and the same equation with $i$ and $j$ interchanged imply that
\begin{equation}\begin{split}\label{eq:}
\psi^\dagger M^{(i,j)} \psi = \delta_{i,j},\quad\psi^\dagger M^{[i,j]} \psi = 0
\end{split}\end{equation}
for any $\psi$. Since the matrices involved are Hermitian, we can choose $\psi$ to be any of their eigenvectors. This forces \emph{all} eigenvalues of $M^{(i,j)}$ to be $\delta_{i,j}$ and those of $M^{[i,j]}$ to be $0$, so that
\begin{equation}
M^{(i,j)} =\delta_{i,j} \openone,\quad M^{[i,j]} = 0,
\end{equation}
where $\openone$ denotes the identity matrix.
It follows that $M^{i,j}=\delta_{i,j} \openone$, i.e.
\begin{equation}
\langle \alpha' ; i | \alpha ; j \rangle =(M^{i,j})_{\alpha',\alpha}= \delta_{i,j} \delta_{\alpha',\alpha}.
\end{equation}

In particular, for any pair of $d_A$-dimensional vectors $\psi$ and $\phi$, we have
\begin{equation}\begin{split}\label{eq:app_Inner}
\langle \psi ; i' | \phi ; i \rangle = \sum_{\alpha',\alpha} \psi_\alpha' \langle \alpha'; i'| \alpha; i \rangle \psi_{\alpha}
= \left( \psi, \phi \right)\, \delta_{i,i'},
\end{split}\end{equation}
where $\left( \psi, \phi \right)$ denotes the usual inner product.
This sharpens our notion of the defect region $A$ ``trapping'' a local Hilbert space of dimension $ d_A $: If there is only one Schmidt state (single $s_i=1$), the ``factorization'' $\mathcal H_A \otimes \mathcal H_B$ is literal, and such inner-product structure is automatic. The above computation shows that in the presence of entanglement in $|0\rangle$, as is in the general case, each ``channel," labeled by $i$,  acts as if the factorization holds, and as we will see this allows us to establish a notion of locality in $\mathcal H_{\rm GS}'$.

Our next task is to expose the entanglement structure within $\mathcal H_{\rm GS}'$. So far, we have only focused on states of the form $| \psi, \varphi\rangle$, which are the analog of ``product states'' between the defect regions (despite, as we have discussed, the many-body state still generally possesses nonzero entanglement). A general state in $ \mathcal H_{\rm GS}'$ would be more entangled. 
Intuitively, we expect that the entanglement should have two components: (i) that already present in $|0\rangle$, and (ii) additional contribution from entangling the defect states.
In the remaining of this subsection we quantify this intuition.

Let $|\Psi\rangle\in \mathcal H_{\rm GS}'$  be defined by
\begin{equation}\begin{split}\label{eq:}
|\Psi\rangle \equiv  \sum_{\alpha=1}^{d_A} \sum_{\beta = 1}^{d_B} \Psi_{\alpha,\beta} |\alpha,\beta\rangle,
\end{split}\end{equation}
where normalization is given by $\sum_{\alpha,\beta} \Psi_{\alpha,\beta}^* \Psi_{\alpha,\beta} = 1$.
We consider again the Schmidt decomposition of $|\Psi\rangle$ with respect to the region $\mathcal R$, and we will show that the entanglement spectrum of $|\Psi\rangle$ factorizes as $\{ s_i\} \otimes \{ \sigma_j\}$, where $\{ s_i\}$ is just that inherited from $|0\rangle$, and $\{\sigma_j\}$  comes from the entanglement in $\Psi_{\alpha,\beta}$.

To see this, consider the singular-value decomposition of $\Psi_{\alpha,\beta}$:
\begin{equation}\begin{split}\label{eq:}
\Psi_{\alpha,\beta} = \sum_{j} \sigma_j W_{\alpha,j} V_{\beta,j}^*,
\end{split}\end{equation}
where $\sum_{\alpha} W_{\alpha,j}^* W_{\alpha,j'} = \sum_{\beta} V_{\beta,j}^* V_{\beta,j'} = \delta_{j,j'}$. 
Note that, for each $j$, $W_{j}$ and $V^*_{j}$ are respectively $d_A$ and $d_B$ dimensional vectors.
As such, we see that
\begin{equation}\begin{split}\label{eq:}
|\Psi\rangle =& \sum_{j,\alpha,\beta} \sigma_j W_{\alpha,j} V_{\beta,j}^* | \alpha,\beta \rangle\\
=& \sum_{i,j} s_i \sigma_j  \left( \sum_{\alpha} W_{\alpha,j} |\alpha; i \rangle \right)
 \left( \sum_{\beta} V^*_{\beta,j} |\beta; i \rangle \right)\\
=& \sum_{i,j}  s_i  \sigma_j  \, \left( | \Psi_A; i,j\rangle  \right) \left(| \Psi _B; i,j\rangle \right),
\end{split}\end{equation}
where in going to the second line we again used the observation that all basis states have the same entanglement spectrum $\{ s_i\}$, and $|\Psi_{A,B} ; i,j\rangle$ in the last line are defined as the corresponding parenthesis. Now check that, by Eq.~\eqref{eq:app_Inner},
\begin{equation}\begin{split}\label{eq:}
\langle \Psi_{A} ; i',j'| \Psi_A; i,j\rangle = & \langle W_{j'}; i' | W_j; i \rangle \\
=& ( W_{j'} , W_j ) \, \delta_{i,i'} = \delta_{i,i'} \delta_{j,j'},
\end{split}\end{equation}
and similarly for $\{ | \Psi_B ; i,j\rangle \}$. By the uniqueness of Schmidt decomposition (i.e.~SVD), we see that the Schmidt weights of $|\Psi\rangle$ are 
\begin{equation}\begin{split}\label{eq:}
\{ s_i \sigma_j \} = \{ s_i \} \otimes \{ \sigma_j \}.
\end{split}\end{equation}
This verifies our claim.
In particular, $|\Psi\rangle$ has the same entanglement spectrum as $| \psi, \varphi\rangle$ if and only if $\{\sigma_j\}$ is a singleton, meaning that $\Psi$ is a ``product state'' $\Psi_{\alpha,\beta} =W_{\alpha,1} V_{\beta,1}^*$. 
More generally, the entanglement entropy is given by the sum of that from $\{ s_i\}$ and that from $\{ \sigma_j\}$, and therefore the ``product states'' $| \psi, \varphi\rangle$ are the only states in $\mathcal H_{\rm GS}'$ with minimal entropy.

\subsection{Symmetry operations in $\mathcal H_{\rm GS}'$}
Next, we utilize the discussion above on entanglement structure to constrain the form of symmetry representations in $\mathcal H_{\rm GS}'$, showing that they indeed factorize in the way described in the main text.

Let $\hat g$ be an on-site unitary operator leaving $\mathcal H_{\rm GS}'$ invariant, and let $N_D$ denotes the number of defect regions.
The projection of $\hat g$ into the ground space is a $\left( \prod_{i=1}^{N_D} d_i \right)$-dimensional unitary matrix $\hat g|_{\rm GS}$, with the entries determined by the expansion 
\begin{equation}\begin{split}\label{eq:gDef}
\hat g \ket{ \alpha_1,\dots, \alpha_{N_D}} = \sum_{\beta_i} (\hat g|_{\rm GS})^{\alpha_1,\dots,\alpha_{N_D}}_{\beta_1,\dots,\beta_{N_D}} \ket{\beta_1,\dots, \beta_{N_D}}.
\end{split}\end{equation}
Similar to before, we imagine an entanglement cut isolating a defect region $\mathcal R^{(i)}$, and we compare the entanglement spectrum obtained from the two sides of Eq.~\eqref{eq:gDef}.
As $\hat g$ is on-site, and that $\ket{\beta_1,\dots, \beta_{N_D}}$ can be rotated into $\ket{\alpha_1,\dots, \alpha_{N_D}}$ using the local unitary operators $\hat U^{(i)}_{(\alpha_i,\beta_i)}$, the entanglement spectrum obtained from  $\hat g\ket{\alpha_1,\dots, \alpha_{N_D}}$ and $\ket{\beta_1,\dots, \beta_{N_D}}$ must be identical (up to exponential accuracy controlled by the localization of $\hat U^{(i)}_{(\alpha_i,\beta_i)}$). This implies the matrix $\hat g|_{\rm GS}$ \emph{cannot} generate any entanglement between different defect regions.
From our earlier discussion, the only states in $\mathcal H_{\rm GS}'$ with the same (minimal) entanglement entropy as the basis states are those described by a factorizable $\Psi_{\alpha_1,\dots,\alpha_{N_D}} $.
The form of $\hat g|_{\rm GS}$  is therefore strongly constrained, and in general we can write \cite{UniversalGate}
\begin{equation}\begin{split}\label{eq:gFac}
\hat g|_{\rm GS} =  \left( \bigotimes_{i=1}^{N_D} \hat g|_{\rm GS}^{(i)}\right)  P_{g},
\end{split}\end{equation}
where $ P_g$ is a permutation of the defect regions, and $ \hat g|_{\rm GS}^{(i)}$ is a $d_i$-dimensional unitary matrix corresponding to the rotation of local degenerate states at the defect region $\mathcal R^{(i)}$.
In addition, for an on-site $\hat g$  the permutation $P_g$ must be trivial. To see this, we simply note that the action of $\hat g|_{\rm GS}$ in a defect region $\mathcal R^{(i)}$ cannot depend on that of any other defect regions, as one can independently alter the local degeneracies $d_i$ by creating $g$-symmetric defects.
Therefore, an on-site unitary symmetry $\hat g$ is represented by $\hat g|_{\rm GS} = \bigotimes_{i=1}^{N_D} \hat g|_{\rm GS}^{(i)}$ in the ground space up to an exponentially small correction, i.e.~$\hat g|_{\rm GS}$ also acts locally in each defect region.

More generally, the symmetry $\hat g$ could by itself be spatial and hence involve site permutations. As long as $\hat g$ does not generate entanglement between different defect regions, say for $\hat g$ being a pure site permutation followed by an on-site unitary transformation, the form Eq.~\eqref{eq:gFac} holds, with $ P_{g}$ determined by the permutation of defect regions dictated by $\hat g$.

Finally, we show that when all the defect regions are related to each other by a spatial symmetry $g$, there exists a choice of basis for which $\hat g |_{\rm GS} =  P_{g}$. Let $m$ be the order of $\hat g$, i.e.~$\hat g ^m = \hat 1$, and label the regions according to $P_g (\mathcal R^{(i)}) = \mathcal R^{(i+1)}$ ($\mod m$).
This implies $d_1=d_2=\dots =d_m$, and the matrices in Eq.~\eqref{eq:gFac} satisfy
\begin{equation}\begin{split}\label{eq:gCons}
\hat g|_{\rm GS}^{(m)} \dots \hat g|_{\rm GS}^{(2)}\,\hat g|_{\rm GS}^{(1)} = \openone.
\end{split}\end{equation}
Under a basis change given by a unitary matrix $u^{(i)}$ for the region $\mathcal R^{(i)}$, we have
\begin{equation}\begin{split}\label{eq:}
\hat g|_{\rm GS}^{(i)} \mapsto    u^{(i) } \left( \hat g|_{\rm GS}^{(i)}\right) u^{(i-1) \dagger}.
\end{split}\end{equation}
To achieve our goal, we simply choose
\begin{equation}\begin{split}\label{eq:}
u^{(i)} =  \hat g|_{\rm GS}^{(1)\dagger} \,\hat g|_{\rm GS}^{(2) \dagger} \dots \hat g|_{\rm GS}^{(i) \dagger},
\end{split}\end{equation}
where in particular we have $u^{(m)} = \openone$ by Eq.~\eqref{eq:gCons}.

\section{Toric code
\label{app:Toric}}
The toric code ($\mathbb{Z}_2$ gauge theory) is one of the simplest models with topological order.  In this appendix, we first explain that the topological order of the toric code is, in fact, guaranteed by our no-go theorem (provided we know the model is gapped and does not spontaneously break symmetry). We then derive the representation of $\hat{C}_2'|_{\text{GS}}$ and $\hat{Z}|_{\text{GS}}$ for this particular model, to see concretely how the ``contradiction'' in our argument [discussed in the section ``Derivation of the rotation no-go'' in the main text] is circumvented in the presence of the topological order.
We thank M. Cheng for a discussion on this subject.

\subsection{The model}
In the toric code, a spin-1/2 lives on each bond of the square lattice (Supplementary Fig.~\ref{fig:toric}a).  
The Hamiltonian reads
\begin{equation}
\hat{H}=-J_e\sum_{\text{vertex }v}\prod_{i\in v}\hat{X}_i-J_m\sum_{\text{plaquette }p}\prod_{i\in p}\hat{Z}_i
\end{equation}
($J_e>0$, $J_m>0$).  Since each term contains an even number of $X_i$'s and $Z_i$'s, the Hamiltonian has the internal $G=\mathbb{Z}_2\times\mathbb{Z}_2$ symmetry represented projectively by spin-1/2's. In addition to this internal symmetry, as the simplest example, here we assume the spatial symmetry of wallpaper group No.~2, generated by the $C_2$ rotation and translation symmetries.  

The wallpaper group fixes the positions of spins to be the center of the bonds corresponding to irreducible Wyckoff positions $w=a$ and $b$ of this wallpaper group.  Then this lattice of spins falls into a nontrivial lattice in the lattice homotopy classification so that the system cannot be in a sym-SRE phase. Hence, knowing that the ground state is gapped and does not spontaneously break symmetries, we can conclude that the ground state has a topological order.

If we put the system on a torus by taking the periodic boundary condition specified in Supplementary Fig.~\ref{fig:toric}(a), for example, then the ground states have four-fold degeneracy and they can be distinguished by the values of the Wilson loops $\prod_{i\in \text{horizontal loop}}\hat{X}_i$ and $\prod_{i\in \text{vertical loop}}\hat{X}_i$.

\subsection{Flux insertion}

\begin{figure}[h]
\begin{center}
{\includegraphics[width=0.5 \textwidth]{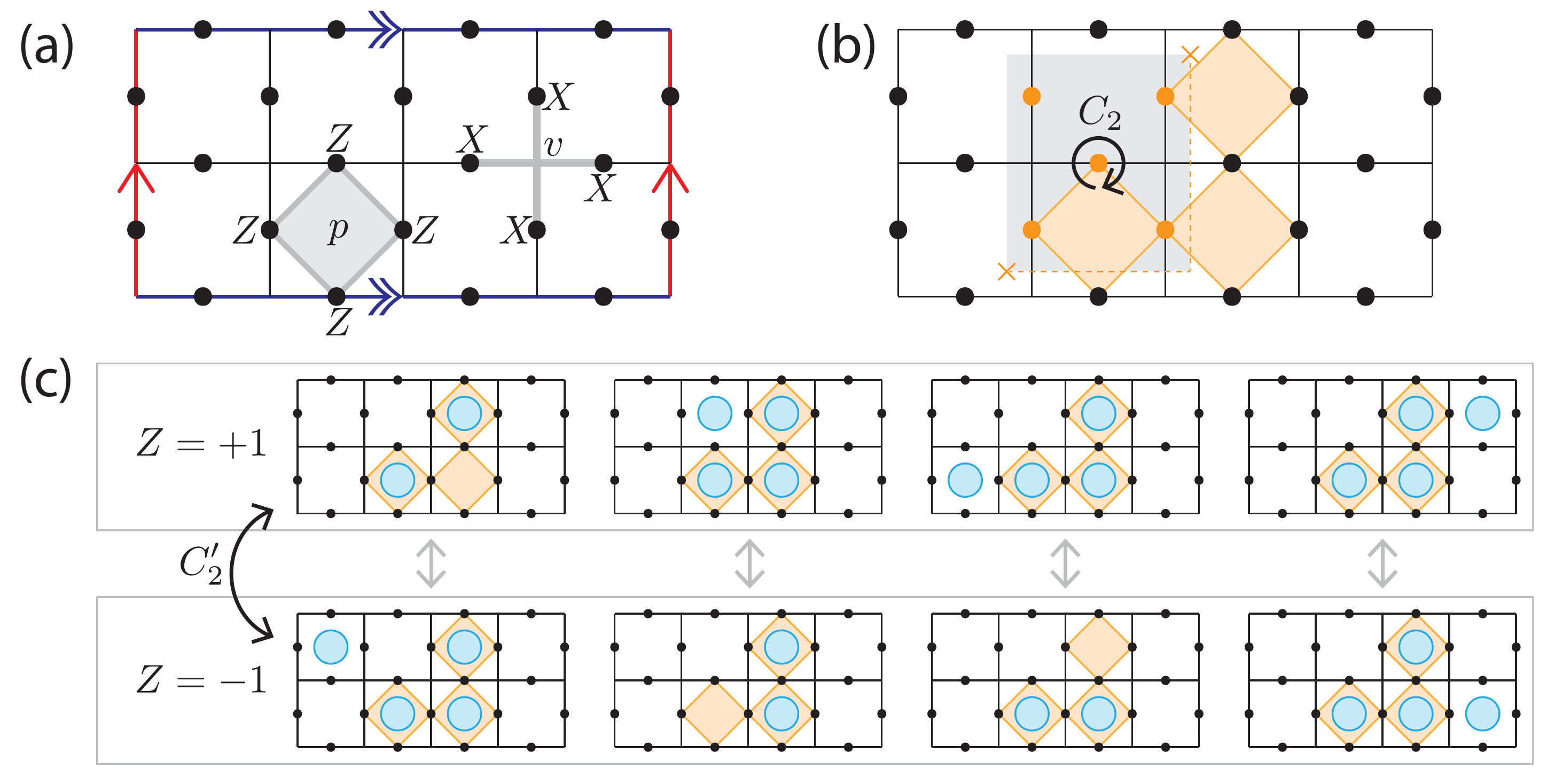}} 
\caption{\label{fig:toric}
(a) The toric code model. Arrows indicate periodic boundary conditions. (b) A configuration of $C_2$-symmetric $X$-fluxes. The orange diamonds represent plaquette flipped by the $X$ fluxes.  (c) Illustration of the ground states after inserting the $X$-fluxes.  The blue circle means that the plaquette has $\prod_{i\in p}\hat{Z}_i=-1$. Those on the top (bottom) row has $+1$ ($-1$) eigenvalue of $\hat{Z}=\prod_{i}\hat{Z}_i$.
 }
\end{center}
\end{figure}

When we insert $X$-fluxes in a $C_2$-symmetric way (Supplementary Fig.~\ref{fig:toric} b), an odd number of plaquette terms flip sign in the Hamiltonian. We marked them by orange diamonds in Supplementary Fig.~\ref{fig:toric}(b). For those plaquettes, $\prod_{i\in p}\hat{Z}_i=-1$ is energetically favored. However, because of the global constraints $\prod_{\text{all plaquettes}}\prod_{i\in p}\hat{Z}_i=+1$, there must be one ``unhappy'' plaquette somewhere in the system in the ground state, and there will be a ground state degeneracy originating from the choice of the position of the unhappy plaquette (Supplementary Fig.~\ref{fig:toric}c) in addition to the four fold degeneracy distinguished by two Wilson loops. This degeneracy is extensive for the model Hamiltonian, but with generic symmetry-preserving perturbations the degeneracy will be reduced, though must preserve at least a two-fold degeneracy due the global anti-commutation relation between $C_2'$ and $\hat{Z}$, as discussed in the main text.
Roughly speaking, the two remaining states correspond to whether the unhappy plaquette is localized near one, or the other, $X$-flux, with the degeneracy between them guaranteed by $C_2'$.
Such an implementation of the degeneracy is precisely the opposite of degeneracy localization, since the two states are related by acting with string-like operator which transports an emergent magnetic flux (i.e., an unhappy plaquette) between the two $X$-flux. 

To see how the global commutation relation is then implemented, recall the flux-inserted Hamiltonian is still symmetric under $\hat{C}_2'=(\prod_{i\in A}\hat{X}_i)\hat{C}_2$.  A pair of ground states interchange under $\hat{C}_2'$ (Supplementary Fig.~\ref{fig:toric}c ). The two paired ground states have the opposite eigenvalue of $\hat{Z}=\prod_{i}\hat{Z}_i$.   Therefore, $\hat{C}_2'$ and $\hat{Z}$ are respectively represented by $\sigma_1$ and $\sigma_3$ within each pair, and thus
\begin{eqnarray}
\hat{C}_2'|_{\text{GS}}&=&\oplus_{\text{all pairs of GS}}\,\,\sigma_1,\\
\hat{Z}|_{\text{GS}}&=&\oplus_{\text{all pairs of GS}}\,\,\sigma_3.
\end{eqnarray}
As required, $\hat{C}_2'|_{\text{GS}}$ and $\hat{Z}|_{\text{GS}}$ do anticommute. However, there is no contradiction in this case because $\hat{C}_2'|_{\text{GS}}$ is not a SWAP operator. To see this, note that the trace of the SWAP operator for localized degeneracies $\{|\alpha_1,\alpha_2\rangle: \alpha_1,\alpha_2=1,\ldots,d\}$ must be $d$, while the $\hat{C}_2'|_{\text{GS}}$ derived here is traceless.  This means that the degeneracy localization assumption is violated, as expected for long-range entangled phases.

\section{Layer Groups
\label{app:Layer}}
In this section we explain that our 2D argument for wallpaper groups actually covers all 80 layer groups.  Layer groups are symmetries of 2D lattices embedded in 3D. Each element $g\in \mathcal{S}$ maps $(x,y,z)$ to $(x',y',z')$, where $(x',y')=p_g(x,y)+\vec{t}_g$ ($p_g$ is a $O(2)$ matrix and $\vec{t}_g$ is a two component vector) and $z'=+z$ or $-z$ depending on $g$.  In particular, the translation subgroup of $\mathcal{S}$ is only in the 2D plane.  When one talks about spinful electrons with significant spin-orbit couplings, it is important to use layer groups instead of wallpaper groups, since it is \emph{not} sufficient to know how $x$ and $y$ transform to identify the transformation of the spin degree of freedom, as (the physical electron) spins are a projective representation of $O(3)$, not $O(2)$.  

In this paper, however, we assume that the spatial symmetry $\mathcal{S}$ acts only on the spatial coordinate and leaves the spin degree of freedom (more generally, the projective representation of $G$) intact. In this case, one can regard the spatial coordinate $z$ as an additional internal degree of freedom.  If one completely forgets about the transformation of $z$, the 2D part of the transformation $(x',y')=p_g(x,y)+\vec{t}_g$ defines one of the 17 wallpaper groups, which we denote by $\mathcal{S}'$. Supplementary Table~\ref{tabLG} summarizes the correspondence.

Now, let $(x,y,z)$ be a point that belongs to an IWP of $\mathcal{S}$. Then the symmetry orbit of $(x,y,z)$ under $\mathcal{S}$ does not include $(x,y,-z)$, since if this was not the case one could ``reduce" the Wyckoff position by setting $z\rightarrow0$. Moreover, the 2D part of the coordinate $(x,y)$ belongs to an IWP of $\mathcal{S}'$.  Therefore, on each site of the corresponding IWP of $\mathcal{S}'$, the additional internal degree of freedom $z$ is ``frozen," i.e., it can only be either one of $z$ or $-z$, or simply $0$, and, after all, it is not really a ``degree of freedom" per se. Therefore one can always set $z=0$, which makes the correspondence between $\mathcal{S}$ and $\mathcal{S}'$ clearer.

\begin{table}[h] 
\begin{center}
\caption{Correspondence between layer groups and wallpaper groups. The layer groups in bold face act trivially on $z$.  The asterisk indicates that, if the transformation of $z$ is completely neglected, the translation subgroup of the layer group is enhanced.\label{tabLG}}
\begin{tabular}{ccccc}\hline\hline
\hspace{5pt}Wallpaper group\hspace{5pt} &\hspace{45pt}Layer group\hspace{45pt} \\ \hline
1 & \textbf{1}, 4, 5$^\ast$\\
2 & 2, \textbf{3}, 6, 7$^\ast$ \\
3 & 8, \textbf{11}, 27, 28$^\ast$, 30$^\ast$, 31$^\ast$, 36$^\ast$\\
4 & 9, \textbf{12}, 29, 33$^\ast$ \\
5 & 10, \textbf{13}, 32$^\ast$, 34$^\ast$, 35\\
6 & 14, 19, \textbf{23}, 37, 38$^\ast$, 41$^\ast$, 48$^\ast$\\
7 & 15, 16, 20, \textbf{24}, 40, 43$^\ast$, 45$^\ast$\\
8 & 17, 21, \textbf{25}, 44\\
9 & 18, 22, \textbf{26}, 47, 39$^\ast$, 42$^\ast$, 46$^\ast$\\
10 & \textbf{49}, 50, 51, 52$^\ast$\\
11 & 53, \textbf{55}, 57, 59, 61, 62$^\ast$, 64$^\ast$\\
12 & 54, 58, 60, \textbf{56}, 63\\
13 & \textbf{65}, 74\\
14 & 67, \textbf{69}, 78\\
15 & 68, \textbf{70}, 79\\
16 & 66, \textbf{73}, 75\\
17 & 71, 72, 76, \textbf{77}, 80\\\hline\hline
\end{tabular}
\end{center}
\end{table}

In this way, one can reduce the problem of a layer group $\mathcal{S}$ to that of the corresponding wallpaper group $\mathcal{S}'$.  The lattice homotopy classification and the proof of no-gos for $\mathcal{S}$ are all identical to those for $\mathcal{S}'$.

\section{The conjecture in 3D
\label{app:3D}}
Here, we present some examples of space groups which demonstrate how our current set of no-gos applies to 3D, and also illustrate how they are insufficient to prove the conjecture in general.

The first example is space group \sg{16} ($P222$), generated by three orthogonal $\pi$-rotations $C_{2,\alpha}$ ($\alpha=x,y,z$) and lattice translations. The space group has eight IWPs: $\vec{r}=(r_x, r_y, r_z)$ where $r_\alpha=0$ or $1/2$~\cite{ITC}. 
For an internal symmetry group $G$ giving $\mathcal H^2(G,{\rm U}(1)) = \mathbb Z_n$, the lattice homotopy can be readily computed and the result is $\mathbb{Z}_n\times(\mathbb{Z}_{\mathrm{gcd}(n,2)})^7$.  Suppose $n=2$, which gives $2^8 = 256$ distinct lattices under lattice homotopy. Whenever the net representation on the intersection of a $C_2$ rotation axis and the primitive unit cell is projective, the ``rotation no-go'' is applicable and therefore this nontrivial lattice (in the lattice homotopy sense) indeed forbids any sym-SRE phase. 
To see this, consider putting the system on a large but finite torus with an odd circumference along the $C_2$ axis of interest, which we suppose is along $\hat z$. We can then formally view the system as two-dimensional, where each ``point'' is interpreted as a loop in the $z$-direction. Our 2D argument will then go through, except that the fluxes are now promoted to a flux loops and the defect line is promoted to a defect surface (with the topology of a cylinder).

On the other hand, when every IWP is occupied by the nontrivial projective representation $[\omega]_{\vec{r}}=-1$, we are unable to prove an obstruction for sym-SRE phases using the no-gos we derived, although this configuration is a nontrivial element in the lattice homotopy classification. In fact, the derived no-gos rule out sym-SRE phases in all but one of the $255$ nontrivial lattices. This also illustrates the fact that, despite we are unable to prove the full conjecture in 3D, our results can still be applied to many 3D lattices, or even to realistic 3D materials. (More precisely, if a lattice is nontrivial under the ``no-go'' classification  discussed in Sec.\ \ref{app:LH=NG}, defined using the three no-gos we derived, a sym-SRE phase is forbidden.)

As an other example, we consider the space group \sg{2} ($P\bar{1}$). This space group has the inversion symmetry in addition to lattice translations.  The IWPs and the lattice homotopy classifications are identical to that of \sg{16}, and the presence/ absence of no-gos also follows from the preceding discussion. Note that, however, in applying the dimension reduction described above, the $z$-coordinate formally becomes an internal degree of freedom, which is ``flipped'' under the action of the 3D inversion. As discussed in Sec.\ \ref{app:Layer}, insofar as the symmetry group is $\mathcal S\times G$, where $\mathcal S$ acts only on the spatial coordinates, this does not affect the arguments.

\section{New no-gos in 3D
\label{app:DNA}}
While we have not yet derived the full set of sym-SRE no-gos required to prove the conjecture in 3D, we discuss here an interesting extension of the no-go arguments. In particular, this discussion can shed some light on cases for which the Bieberbach argument presented in Supplementary Ref.~\cite{PNAS} does not lead to a provably-tight bound on the filling condition.

We begin by reviewing the ``tight no-go problem'' described in Supplementary Ref.~\cite{PNAS}, which studied a more general setting for which the electrons can be delocalized and spin-orbit coupled, assuming instead only time-reversal symmetry. Generally, a sym-SRE no-go detects some, but not all, obstructions to realizing sym-SRE phases in a system. For instance, consider the original Lieb-Schultz-Mattis theorem, which states that a sym-SRE phase is forbidden whenever $\nu$, the electron filling per primitive unit cell, satisfies $\nu \not \in 2 \mathbb N$. This implies sym-SRE phases are not forbidden when $\nu$ is even, but depending on the symmetries of the systems there can be further no-gos that are not detected by this single criterion. We say a set of no-go is ``tight'' if a sym-SRE phase is possible whenever the no-gos are silent. Proving tightness is generally nontrivial, since it involves explicit constructions of sym-SRE phases for all instances permitted by the no-gos. 

In Supplementary Refs.~\cite{PNAS}, the tightness of the Bieberbach no-go is only proven for 218 of  the 230 space groups. 
Such tightness was established by studying the possible fillings of band insulators \cite{SA,PRL}.
The 12 exceptional SGs correspond to cases where at certain fillings band insulators are disallowed despite the Bieberbach no-go being silent. (As a technical remark, we note the number of exceptional SGs depends on whether spin-orbit coupling and / or TR are assumed or not, as the set of band-insulator fillings can be modified by the symmetry settings \cite{PRL}. Here, ``12'' refers to case of TR symmetric system with negligible spin-orbit coupling.)

The tightness of the Bieberbach no-gos for these 12 exceptional SGs is still an open question. Here, we study some aspect of  this problem by extending the new no-gos developed in the present work to these SGs. In particular, recall that here we are focusing on localized spin problems with on-site unitary symmetries in the limit of vanishing spin-orbit coupling. This is a much more restrictive setting than the one considered in Supplementary Ref.~\cite{PNAS}, and as such it may not be too surprising if one can expose further obstructions, as we will now demonstrate.
	
For concreteness, we focus on only one of the 12 exceptional SGs, {\bf \em{106}}, and consider the most physically-relevant setting for which the internal-symmetry group is taken to be $G = {\rm SO}(3)$, appropriate for crystals with negligible spin-orbit coupling. The Bieberbach no-go forbids any sym-SRE phase unless the electron filling $\nu \in 4 \mathbb N$ \cite{PNAS}, whereas band  insulators are possible only for $\nu \in 8 \mathbb N$ \cite{PRL}. The mismatch between the two, $\nu = 8 n- 4$ for any $n\in \mathbb N$, corresponds to cases for which the tightness of the interacting bounds is unclear. Here, we show that within the class of localized spin models we considered here, a no-go is in fact present and obstructs any sym-SRE phases when $\nu =8n - 4$.

To proceed, we first discuss the symmetries of SG {\bf \em{106}}. The SG is primitive tetragonal, 
and we take the primitive lattice vectors as $\vec t_1 = (1,0,0)$, $\vec t_2=(0,1,0)$ and $\vec t_3 =(0,0,1)$. It can be viewed as being generated by the following two nonsymmorphic symmetries together with the lattice translations:
\begin{equation}\begin{split}\label{eq:}
S_{z} = T_3^{1/2}R_{z,\pi/2};~~G_x= T_1^{1/2} T_2^{1/2}m_y,
\end{split}\end{equation}
where $R_{z,\theta}$ denotes a rotation by angle $\theta$ about the positive $z$ axis, $m_y$ denotes the mirror $y \mapsto -y$, and $T_i$ is the lattice translation by $\vec t_i$. Note that $S_{z}$ is a $4_2$ screw, meaning that 
$ S_z^2 = T_3  R_{z,\pi}$ is a symmorphic $C_2$ rotation. While the point group is of order $8$, sites can sit on the $C_2$ rotation axis and hence one can find higher-symmetry Wyckoff positions with only $4$ sites per unit cell. 
This can be seen from Supplementary Ref.~\cite{ITC}, which lists three Wyckoff positions with $|\mathcal W^\text{\bf \em{106}}_{\rm a}|=|\mathcal W^\text{\bf \em{106}}_{\rm b}| =4$ and the generic position $|\mathcal W^\text{\bf \em{106}}_{\rm c}| = 8$. Both $\mathcal W^\text{\bf \em{106}}_{\rm a}$ and $\mathcal W^\text{\bf \em{106}}_{\rm b}$ are irreducible, and from a lattice-homotopy point of view the lattices are classified by $\mathbb{Z}_2\times \mathbb{Z}_2$ (for $\mathbb{Z}_2$ projective representations).

A general lattice in SG {\bf \em{106}} can be viewed as a stack of $n_{w}$ copies of the minimal lattices in $\mathcal W^\text{\bf \em{106}}_w$ for $w = {\rm a}, {\rm b}$ or ${\rm c}$, and such a lattice will contain $4(n_{\rm a} + n_{\rm b}) + 8 n_{\rm c}$ sites in the primitive unit cell. For strongly-localized electronic systems, each site in the lattice represents a spin-1/2 electron, and therefore the electron filling is simply given by the number of sites. The fillings of interest, $\nu = 8 n - 4$, therefore corresponds to lattices with $n_{\rm a} + n_{\rm b}$ being odd. 

To show that a sym-SRE is forbidden (in this localized-spin, spin-orbit-coupling-free) whenever $\nu =8 n - 4$, it suffices to consider the minimal realizations of lattices in $\mathcal W^\text{\bf \em{106}}_{\rm a}$ and $\mathcal W^\text{\bf \em{106}}_{\rm b}$. First, we consider a minimal lattice in $\mathcal W^\text{\bf \em{106}}_{\rm a}$, which contains four sites in the primitive unit cell with coordinates
\begin{equation}\begin{split}\label{eq:}
\begin{array}{ll}
\vec r_1= (0,0,z);  &\vec r_2 = (0,0,z+1/2);\\
\vec r_3= (1/2,1/2,z); &\vec r_4=(1/2,1/2,z+1/2),
\end{array}
\end{split}\end{equation}
where $z$ is a free parameter. Note that all sites are invariant under a $C_2$ rotation, and are all symmetry-related to each other, e.g.~$S_z(\vec r_1) = \vec r_2$, and $G_x(\vec r_1) = \vec r_3$.

As set up, the Bieberbach no-go is silent. In addition, any $C_2$ axis intersects exactly two sites in each unit cell, and therefore a direct generalization of the $C_2$ no-go to the 3D setting will also be silent. 
The key idea for deriving a no-go is that the $C_2$ no-go can become active if we identify the two $S_z$-related sites in each unit cell sitting on the same rotation axis, i.e.~a no-go is exposed once we combine the Bieberbach ideas with the rotation no-gos.

More precisely, we identify points in $\mathbb R^{3}$ that are related by any elements in the group $\Gamma \equiv \langle \vec t_1, \vec t_2, S_z\rangle$. As $\Gamma$ is fixed-point free, $\mathbb R^3/\Gamma$ is a Bieberbach manifold \cite{PNAS} and we consider the corresponding system defined on this compact space. Note that the volume of $\mathbb R^3/\Gamma$ is only half of the primitive unit cell, which is far away from the thermodynamic limit. This, however, is only a minor technical subtlety, as one can always take a ``scaled-up'' version of $\Gamma$ and obtain a manifold that is as large as one pleases. All the following discussions will be unaffected by such scaling-up.

\begin{figure}
\begin{center}
{\includegraphics[width=0.48 \textwidth]{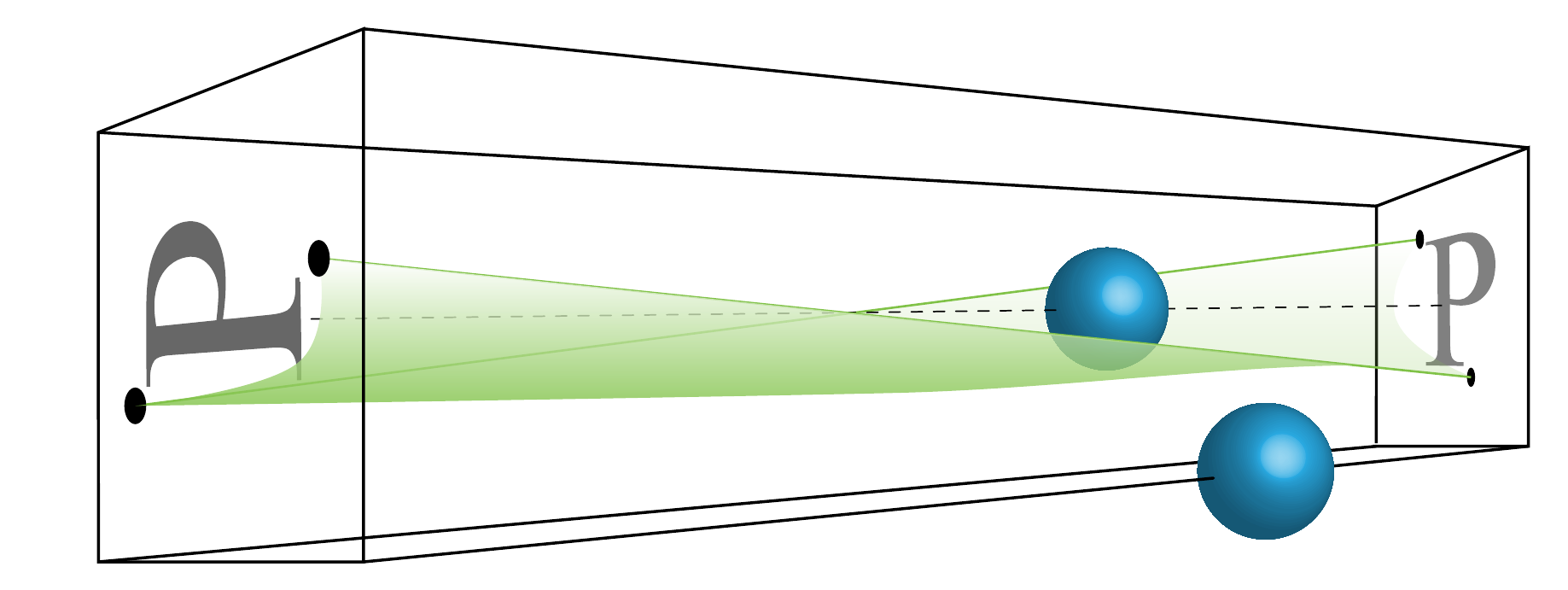}} 
\caption{{\bf New no-go in 3D.} The flux-insertion argument for sym-SRE no-go can be generalized to higher dimensions with more complicated geometries. Here we show a compact space obtained by identifying the two marked faces by a $4_2$ screw and the other pairs by lattice translations. A pair of symmetry-related line fluxes (green lines) are inserted, and the system maintains a $C_2$ rotation symmetry about the dashed axis. The defect surface (shaded) and its $C_2$ partner (not shown) together enclose a single site (sphere).
\label{fig:DNA}}
\end{center}
\end{figure}

Importantly, $C_2$ satisfies the condition listed in Supplementary Ref.~\cite{PRL} and therefore is a ``remnant symmetry,'' i.e.~the modded-out system defined on $\mathbb R^3/\Gamma$ remains $C_2$ symmetric. Now imagine introducing a pair of defect loops in the modding-out procedure, and we consider a topologically-nontrivial configuration for which the loop closes itself with a $S_z$ twist (Supplementary Fig.~\ref{fig:DNA}). In particular, we can position the defect loops such that they form a $C_2$-related pair.
Now we insert $X$-fluxes into the system.
The flux insertion procedure is then similar to the 2D case, except that now we twist the local Hamiltonian along a defect surface bounded by the two defect loops. As can be seen from Supplementary Fig.~\ref{fig:DNA}, under a $C_2$ rotation the defect surface is transformed to its $C_2$ partner, which can be deformed into the original surface by applying a gauge transformation within the enclosed region of space. In the present case, only one site is enclosed, and therefore we have exposed the projective representation carried by the spin-1/2 sitting at that site. The rest of the argument will then proceed in the same way as that for the original rotation no-go.
The only distinction is that the two disconnected defected regions are lines, rather than points, but this makes no difference to the derivation of no-gos (note that we define the system on thermodynamic large, but finite, manifolds).

To complete the argument, we simply note that the above construction applies equally well to the minimal lattice in $\mathcal W^\text{\bf \em{106}}_{\rm b}$ through a shift of origin, and in fact to any lattices with $n_a + n_b$ odd. 
(Actually, we achieved even more, since the argument presented already shows $\sim_{\rm LH} = \sim_{\rm NG}$ for this particular SG.)
This establishes a no-go for any spin-rotation-symmetric quantum magnets in SG {\bf \em{106}} with $\nu = 8 n- 4$. In addition, for each filling $\nu \in 8 \mathbb N$ one can design a symmetry-preserving valence-bond solid with the desired filling. This can be achieved by  putting sites in the generic position and pinning the center of the valence bonds to either $\mathcal W^\text{\bf \em{106}}_{\rm a}$ or $\mathcal W^\text{\bf \em{106}}_{\rm b}$. This implies the tight filling constraints for a sym-SRE phase in SG {\bf \em{106}} is $\nu \in 8 \mathbb N$ in the spin-orbit-free quantum-magnet setting. 

We note that when spin-orbit coupling is negligible and the system is in addition TR invariant, we have established the sym-SRE constraint $\nu \in 8 \mathbb N$ in \emph{both} the free \cite{PRL} and infinite, on-site repulsion limits. If a sym-SRE phase is indeed possible for $\nu = 8n -4$, it will require either changing the symmetry setting, or require interaction strengths that are necessarily ``moderate.''   Somewhat surprisingly, in Supplementary Ref.~\cite{reBS} it was established that if time-reversal is \emph{absent} but spin-rotation symmetry remains intact, it is possible to realize a band insulator at the filling of $\nu=4$. (Note that Supplementary Ref.~\cite{reBS} provides a band insulator example at filling $\nu=2$ for spinless particles. Restoring spin but assuming spin-rotation invariance, this corresponds to a filling $\nu=4$ band insulator.) For such problems, if an on-site (density-density) repulsion $U$ is incorporated, the system necessarily undergoes an interaction-driven phase transition from a sym-SRE phase to either a symmetry broken or a more exotic phase as $U$ is tuned from $0\rightarrow \infty$. It is currently unclear whether the phase transition happens in the spin or charge sector, and it is an interesting open problem to understand the full landscape of filling constraints for this SG in the different symmetry settings.  In closing, we also remark that similar extension of the no-gos apply to some, if not all, of the exceptional SGs, and surveying all of them is another interesting problem that we leave for future studies.

\section{Connection to crystalline symmetry-protected topological orders
\label{app:SPT}}
First, we note our conjecture has a simple interpretation in terms of ``valence bond" patterns, which are commonly used in the discussion of SPTs arising in quantum magnets, as in the pictorial description of the AKLT chain. In this language, our conjecture states that a sym-SRE is possible only when there could be a symmetric valence-bond type state. To see this, consider a lattice of spins $\Lambda$ which is trivial, i.e., $\Lambda \sim 0$, and let $\gamma$ denote the path which annihilates the lattice. 
Formally, we can view $\Lambda$ as an $\mathcal{S}$-invariant 0-chain (a collection of points), and we can think of $\gamma$ as an $\mathcal{S}$-invariant 1-chain (a collection of paths), where the points and lines take values in $\mathbb{Z}_2$.
The statement of homotopy can then be phrased in terms of the boundary of the chain, $\Lambda = \partial \gamma$.
If two sites are connected by a path in $\gamma$, we can view the connection as a valence-bond which projects them into the singlet state. 
The $\mathcal{S}$-invariance of $\gamma$ implies the resulting valence-bond pattern respects the spatial symmetries.

In addition, there is also an intriguing similarity between our conjecture  and recent results on the classification of point-group SPT phases \cite{PhysRevX.7.011020, ThorngrenElse}. In the case of translation symmetry, it was realized that  LSM-constraints in $d$-dimensions  have a deep connection to the physics of SPT phases in $d+1$-dimensions \cite{ChengZaletel}.
The $d$-dimensional boundary of a $d+1$-dimensional SPT must be either symmetry-broken, or LRE  --  just like the LSM-constraint.
It was pointed out that a 2D spin-1/2 magnet can actually be considered as the surface of the 3D AKLT phase, which is an SPT.
Using the SPT bulk-boundary correspondence \cite{Fidkowski2015, WangLinLevin2016}, one can argue not only the usual LSM-constraint, but a generalized constraint on the precise set of topological orders that can emerge. 
Thus it is natural to suspect that our space-group LSM will relate to the classification of  space-group SPTs, a subject of current investigation.
Presumably, each of the lattice equivalence classes $[\Lambda]$ corresponds to a distinct anomaly class of a surface topological order \cite{Fidkowski2015, HermeleChen}.
The lattice homotopy problem we propose is in fact very reminiscent of a recent work which proposes a classification of point-group SPTs in terms of ``packing" and ``stacking" lower dimensional SPTs \cite{PhysRevX.7.011020}.
If crystalline bulk-boundary theorem can be made precise, our current result may follow as a special case.\cite{ElseFuture}

\end{document}